\documentclass[aps,prx,twocolumn,superscriptaddress]{revtex4-1}
\usepackage{graphicx,braket,hyperref,amsmath,comment,float}
\hypersetup{
	colorlinks   = true,
	allcolors    = [rgb]{0,0,1.0}
}

\begin{document}

\title{Selective probing of magnetic order on Tb and Ir sites in stuffed Tb$_{2+x}$Ir$_{2-x}$O$_{7-y}$
using resonant  X-ray scattering}

\author{C.~Donnerer}
\affiliation{London Centre for Nanotechnology and Department of Physics and Astronomy, University College London, London WC1E 6BT, United Kingdom}

\author{M.~C.~Rahn}
\thanks{Present affiliation: Los Alamos National Laboratory, Los Alamos, New~Mexico 87545, USA}
\affiliation{Department of Physics, University of Oxford, Clarendon Laboratory, Oxford, OX1 3PU, United Kingdom}

\author{E.~Schierle}
\affiliation{Helmholtz-Zentrum Berlin f{\"u}r Materialien und Energie, Albert-Einstein-Stra{\ss}e 15, D-12489 Berlin, Germany}

\author{R.~S.~Perry}
\affiliation{London Centre for Nanotechnology and Department of Physics and Astronomy, University College London, London WC1E 6BT, United Kingdom}
\affiliation{ISIS Facility, Rutherford Appleton Laboratory, Chilton, Didcot, OX11 0QX, United Kingdom}

\author{L.~S.~I.~Veiga}
\affiliation{London Centre for Nanotechnology and Department of Physics and Astronomy, University College London, London WC1E 6BT, United Kingdom}

\author{G.~Nisbet}
\affiliation{Diamond Light Source Ltd, Diamond House, Harwell Science and Innovation Campus, Didcot, Oxfordshire OX11 0DE, United Kingdom}

\author{S.~P.~Collins}
\affiliation{Diamond Light Source Ltd, Diamond House, Harwell Science and Innovation Campus, Didcot, Oxfordshire OX11 0DE, United Kingdom}

\author{D.~Prabhakaran}
\affiliation{Department of Physics, University of Oxford, Clarendon Laboratory, Oxford, OX1 3PU, United Kingdom}

\author{A.~T.~Boothroyd}
\affiliation{Department of Physics, University of Oxford, Clarendon Laboratory, Oxford, OX1 3PU, United Kingdom}

\author{D.~F.~McMorrow}
\affiliation{London Centre for Nanotechnology and Department of Physics and Astronomy, University College London, London WC1E 6BT, United Kingdom}

\begin{abstract}
We study the magnetic structure of the ``stuffed" (Tb-rich) pyrochlore iridate Tb$_{2+x}$Ir$_{2-x}$O$_{7-y}$, using resonant elastic x-ray scattering (REXS). 
In order to disentangle contributions from Tb and Ir magnetic sublattices, experiments were performed at the Ir $L_3$ and Tb $M_5$ edges, which provide selective sensitivity to Ir $5d$ and Tb $4f$ magnetic moments, respectively.
At the Ir $L_3$ edge, we found the onset of long-range ${\bf k}={\bf 0}$ magnetic order below $T_{N}^\text{Ir}\sim$\,71\,K, consistent with the expected signal of all-in all-out (AIAO) magnetic order. 
Using a single-ion model to calculate REXS cross-sections, we estimate an ordered magnetic moment of $\mu_{5d}^{\text{Ir}} \approx 0.34(3)\,\mu_B$ at 5\,K. 
At the Tb $M_5$ edge, long-range ${\bf k}={\bf 0}$ magnetic order appeared below $\sim40\,$K, also consistent with an AIAO magnetic structure on the Tb site.
Additional insight into the magnetism of the Tb sublattice is gleaned from measurements at the $M_5$ edge in applied magnetic fields up to 6\,T, which is found to completely suppress the Tb AIAO magnetic order.
In zero applied field, the observed gradual onset of the Tb sublattice magnetisation with temperature suggests that it is induced by the magnetic order on the Ir site. 
The persistence of AIAO magnetic order, despite the greatly reduced ordering temperature and moment size compared to stoichiometric Tb$_{2}$Ir$_{2}$O$_{7}$, for which $T_{N}^{\text{Ir}} =130\,$K and $\mu_{5d}^{\text{Ir}}=0.56\,\mu_B$, indicates that stuffing could be a viable means of tuning the strength of electronic correlations, thereby potentially offering a new strategy to achieve topologically non-trivial band crossings in pyrochlore iridates.
\end{abstract}

\maketitle

\section{Introduction}

The exploration of non-trivial band topology in the presence of electronic correlations offers a promising route to discover novel phases of matter~\cite{krempa2014correlated, schaffer2016recent}. 
Following the seminal proposal of a Weyl semimetal (WSM) phase in magnetically ordered pyrochlore iridates $R_2$Ir$_2$O$_7$ (where $R$ is a rare-earth element), these materials have attracted special interest in this search~\cite{wan2011topological, krempa2012topological, go2012correlation, krempa2013pyrochlore}.
These studies acted as a catalyst in this research area, and a number of uncorrelated WSM states have since been discovered, e.g. in monopnictides~\cite{lv2015experimental, lv2015observation, xu2015discovery1, xu2015discovery2, yang2015weyl}. Nevertheless, it remains unclear whether or not the strongly correlated $d$ electronic states in pyrochlore iridates can host similar topologically protected band crossings~\cite{shinaoka2015phase, zhang2017metal-insulator}.

Early theoretical work on pyrochlore iridates suggested that two conditions are required for a WSM phase: (1)~The Ir magnetic moments need to order in an AIAO fashion (where all moments point towards or away from the centre of the Ir tetrahedra in the pyrochlore lattice), thereby preserving the overall cubic symmetry while breaking that of time reversal; and (2)~the electronic correlations must be weak enough to prevent a more conventional Mott insulating phase~\cite{wan2011topological,krempa2012topological, krempa2014correlated, yang2014emergent, shinaoka2015phase, zhang2017metal-insulator}. Many studies have set out to address condition (1), and empirical evidence for AIAO magnetic order has been found in several pyrochlore iridates, using a variety of techniques, including muon spin rotation ($\mu$SR)~\cite{zhao2011magnetic, disseler2012magnetic1, disseler2012magnetic2, graf2014magnetism}, neutron powder diffraction (NPD)~\cite{tomiyasu2012emergence, lefrancois2015anisotropy, guo2016direct, guo2017magnetic} and single-crystal REXS~\cite{sagayama2013determination, clancy2016x-ray, donnerer2016all-in}. However, the second condition of weak electronic correlations has been under some dispute: several theoretical and experimental studies have challenged the assumptions made by early density functional theory and mean-field calculations, and instead argued that pyrochlore iridates are strongly correlated across the rare-earth series~\cite{shinaoka2015phase, ueda2016variation, zhang2017metal-insulator}. 

A key challenge in the synthesis of pyrochlore iridates, and pyrochlore oxides in general, is that small changes in stoichiometry are common and can have substantive effects on electronic and magnetic properties~\cite{gardner2010magnetic, MacLaughlin2015unstable, Telang2018dilute}. In particular, flux-grown pyrochlore iridate crystals tend to exhibit some degree of ``stuffing", where an excess of $R$ ions populate the Ir sites, leading to a $R_{2+x}\text{Ir}_{2-x}\text{O}_{7-y}$ stoichiometry. Studies have shown that small values of $x$ can suppress the onset of magnetic order on the Ir site, and lead to a more metallic state~\cite{nakayama2016slater}.
At some critical value of $x$, the magnetic instability in pyrochlore iridates is fully suppressed~\cite{graf2014magnetism, clancy2016x-ray}.
While this effect has so far largely been considered an experimental nuisance \cite{Telang2018dilute}, it may offer an interesting way to achieve a more weakly correlated state, and hence provide a route to stabilise topological phases. However, the fate of the magnetic order in the presence of stuffing has not yet been thoroughly studied in pyrochlore iridates.

In this study, we show that it is possible to significantly suppress the onset of magnetic order in stuffed Tb$_{2+x}$Ir$_{2-x}$O$_{7-y}$, from the stoichiometric value of $T_{N} = 130\,$K \cite{matsuhira2011metal} to about $71\,$K, while retaining the AIAO magnetic structure. Furthermore, by performing REXS experiments at both Tb $M_5$ and Ir $L_3$ edges, we disentangle the magnetic contributions of Tb and Ir sites, confirming the view that the magnetic order on the Ir site induces magnetic order on the Tb site~\cite{lefrancois2015anisotropy}. Finally, using single-ion calculations of REXS cross-sections, we show that the size of the Ir ordered magnetic moment is strongly reduced, to $0.34(3)\,\mu_{B}$ at $5\,$K. Our results show that stuffing of pyrochlore iridates provides a means to tune the electronic groundstate without destroying magnetic order, which may aid the realisation of novel topological states.

\section{Experimental Methods}

Tb$_{2+x}$Ir$_{2-x}$O$_{7-y}$ single crystals were flux-grown at the Clarendon Laboratory (Oxford, UK), as described in Ref.~\cite{Millican2007crystal}. They were characterised by superconducting quantum interference device (SQUID) magnetometry, four-probe resistance measurements and laboratory X-ray diffraction (XRD). Energy dispersive X-ray spectroscopy (EDX) was used to investigate the cation ratio, which revealed an excess of Tb ions of $x \sim 0.18$. This excess of Tb ions is also evident in the somewhat enlarged lattice parameter at $300\,$K, in our sample we measured $a = 10.25\,$\AA{}, compared to a literature value of $a = 10.23\,$\AA{}~\cite{lefrancois2015anisotropy}. No reliable estimate of the oxygen content of our single crystals could be obtained.

REXS experiments at the Ir $L_3$ edge (dominated by dipolar $2p \rightarrow 5d$ transitions at a photon energy of 11.215\,keV) of Tb$_{2+x}$Ir$_{2-x}$O$_{7-y}$ were performed at beamline I16 of the Diamond Light Source (Didcot, UK). A vertical scattering geometry was used, the incident polarisation provided by an in vacuum U27 undulator was linear, and perpendicular to the scattering plane ($\sigma$ polarisation). The polarisation of the scattered beam was analysed with a Au (333) crystal. An orientated Tb$_{2+x}$Ir$_{2-x}$O$_{7-y}$ single crystal was mounted on the cold finger of a closed-cycle refrigerator, such that (100) and (011) directions were in the vertical scattering plane.

REXS experiments at the Tb $M_5$ edge ($3d \rightarrow 4f$, photon energy of 1.241\,keV) of Tb$_{2+x}$Ir$_{2-x}$O$_{7-y}$ were performed at beamline UE46-PGM-1 of BESSY II (Berlin, Germany). This beamline operates in the soft X-ray spectrum (120 eV and 2000 eV) using radiation from an elliptical undulator, which allows the incident linear polarisation to be rotated from horizontal ($\pi$) to vertical ($\sigma$). A horizontal scattering geometry was used, and the same Tb$_{2+x}$Ir$_{2-x}$O$_{7-y}$ sample was mounted on a He-flow cryostat, with the (100) and (011) directions in the scattering plane.

\section{Results}

\subsection{Bulk properties}

\begin{figure}[htb]
\centering
\includegraphics[width=5cm]{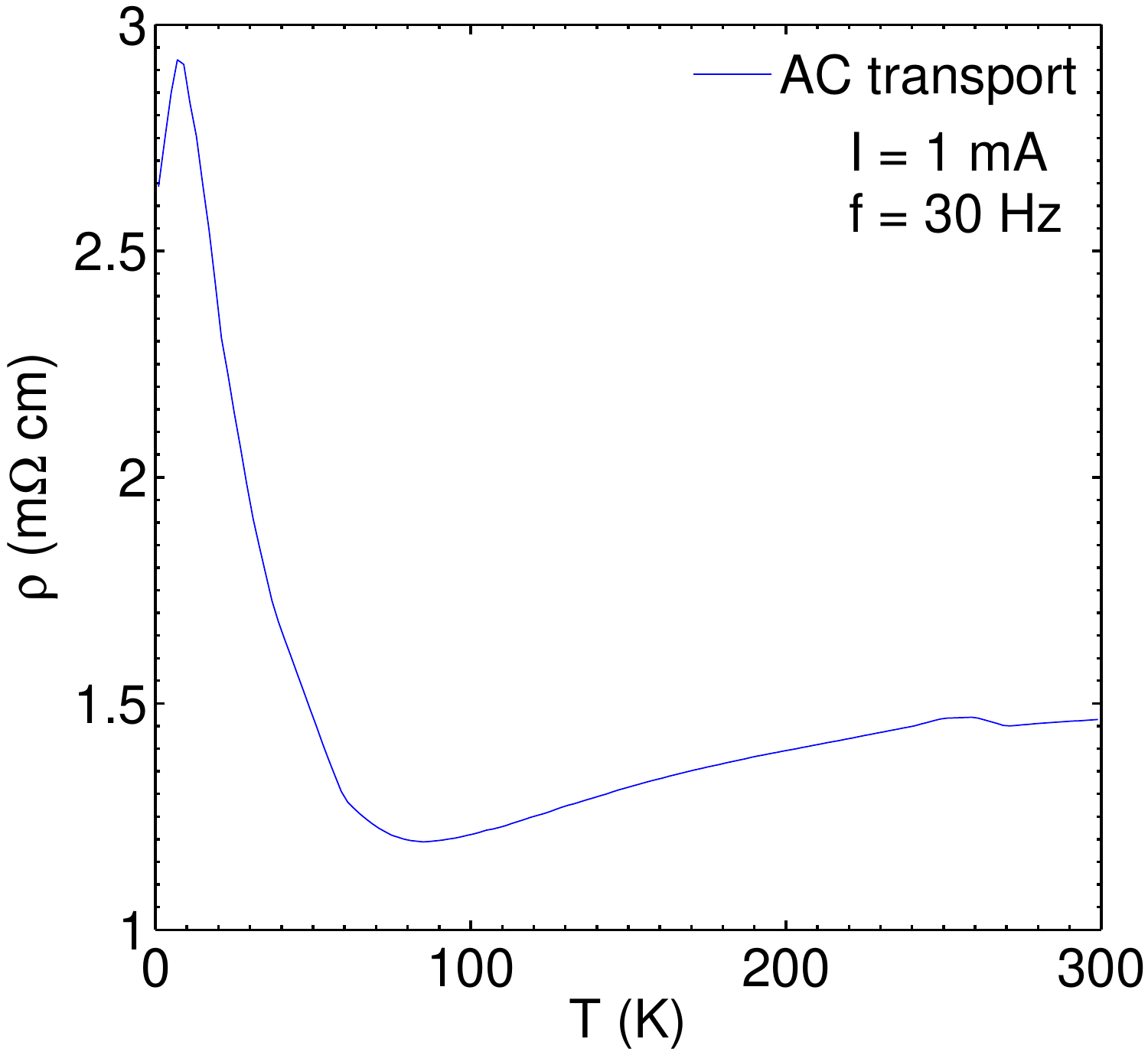}\\
\includegraphics[width=5.0cm]{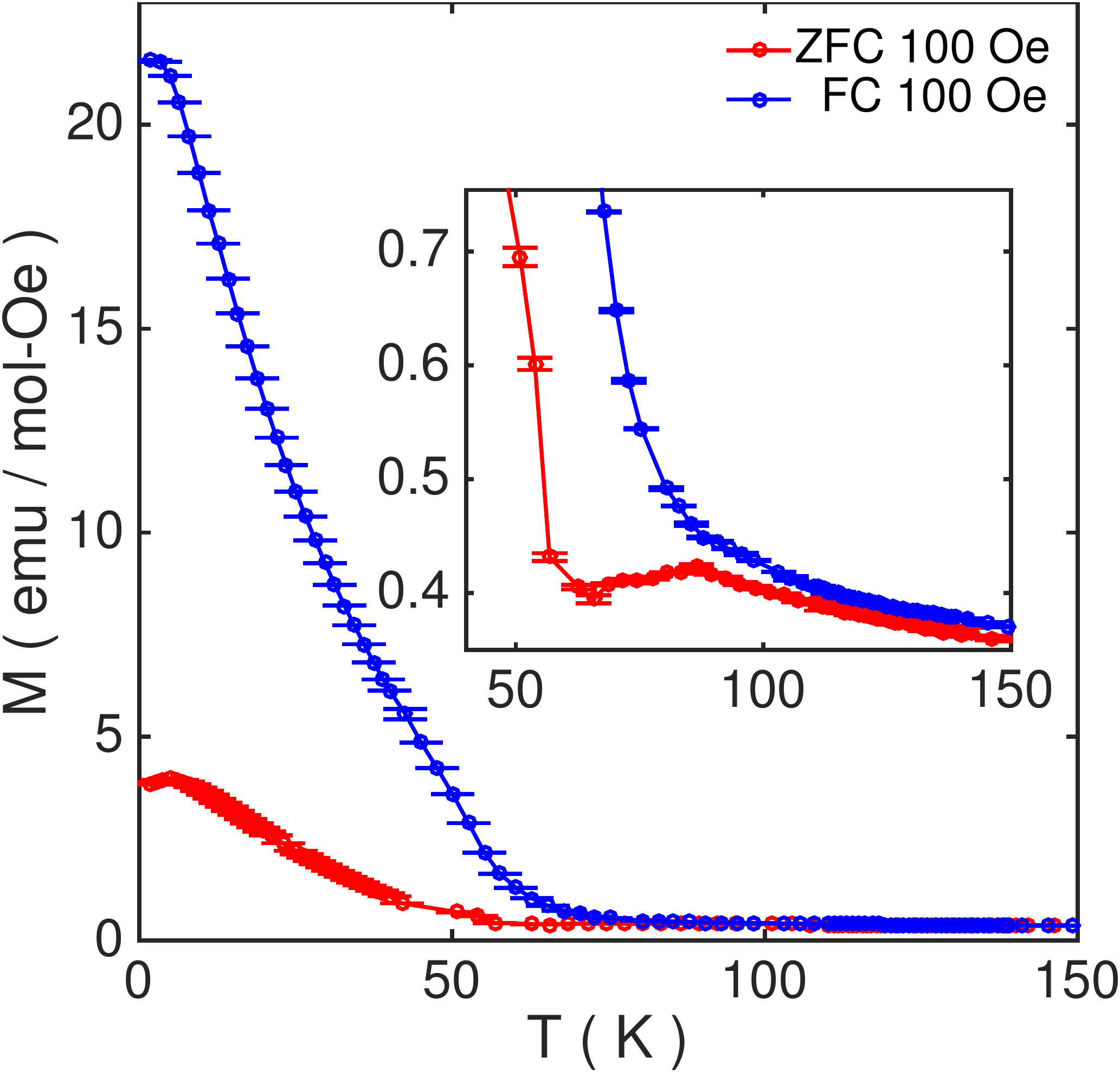}
\caption{(Top) Resistivity of Tb$_{2+x}$Ir$_{2-x}$O$_{7-y}$ single-crystal as a function of temperature. (Bottom) Temperature dependence of the zero field cooled (ZFC) and field cooled (FC) magnetisation of Tb$_{2+x}$Ir$_{2-x}$O$_{7-y}$ single-crystal.}
\label{Tb227_bulk}
\end{figure}

For nominally stoichiometric powder samples of Tb$_{2}$Ir$_{2}$O$_{7}$, published magnetisation data suggest that the Ir site orders at $\sim130\,$\,K~\cite{matsuhira2011metal, lefrancois2015anisotropy}, while resistivity data show insulating behaviour at all temperatures, with a small anomaly at the magnetic transition~\cite{matsuhira2011metal}.
The magnetic structure of Tb$_{2}$Ir$_{2}$O$_{7}$ has been studied by neutron powder diffraction~\cite{lefrancois2015anisotropy, guo2017magnetic}. While lacking the element specificity of the REXS technique, the NPD experiments indicate that both the Tb and Ir sublattices exhibit AIAO order below a N\'eel temperature of $\sim125$--$130$\,K. One notable feature of the NPD experiments is that the onset of order on the Tb sublattice is more gradual than that of the Ir one, suggesting that the former is induced by spontaneous order on the latter, through significant  $d$--$f$ interactions. The low-temperature (1.5\,K) values of the magnetic moments were found to be $5.24\,\mu^{}_B$ and $0.56\,\mu^{}_B$ for the Tb and Ir sites, respectively~\cite{guo2017magnetic}.

Figure \ref{Tb227_bulk} shows the magnetisation and resistivity data as a function of temperature for our single crystal sample of Tb$_{2+x}$Ir$_{2-x}$O$_{7-y}$. At about $70\,$K, a bifurcation in the field-cooled (FC) and zero-field-cooled (ZFC) magnetisations is observed, which indicates that the N\'{e}el temperature is strongly reduced from its literature value of $\sim130\,$K. Furthermore, the resistivity appears significantly more metallic, with absolute values 2--3 orders of magnitude lower than the ones reported in powder samples~\cite{matsuhira2011metal}. The discrepancies between our single crystal sample and published data on powders of Tb$_2$Ir$_2$O$_7$ most likely arise from excess Tb ions populating Ir sites, resulting in an stoichiometry of Tb$_{2+x}$Ir$_{2-x}$O$_{7-y}$, with $x\sim 0.18$ as revealed by our EDX experiments. We note that annealing the Tb$_{2+x}$Ir$_{2-x}$O$_{7-y}$ sample in an oxygen atmosphere did not alter its bulk properties.

At around 10\,K, anomalies occur in both the magnetisation and resistivity, similar to observations in the literature~\cite{lefrancois2015anisotropy}. These have been argued to be associated with $f$--$f$ interactions between magnetic Tb ions.

\subsection{REXS at the Ir $L_3$ edge}

\subsubsection{Magnetic structure}

\begin{figure}[thb]
\centering
\includegraphics[width=\linewidth]{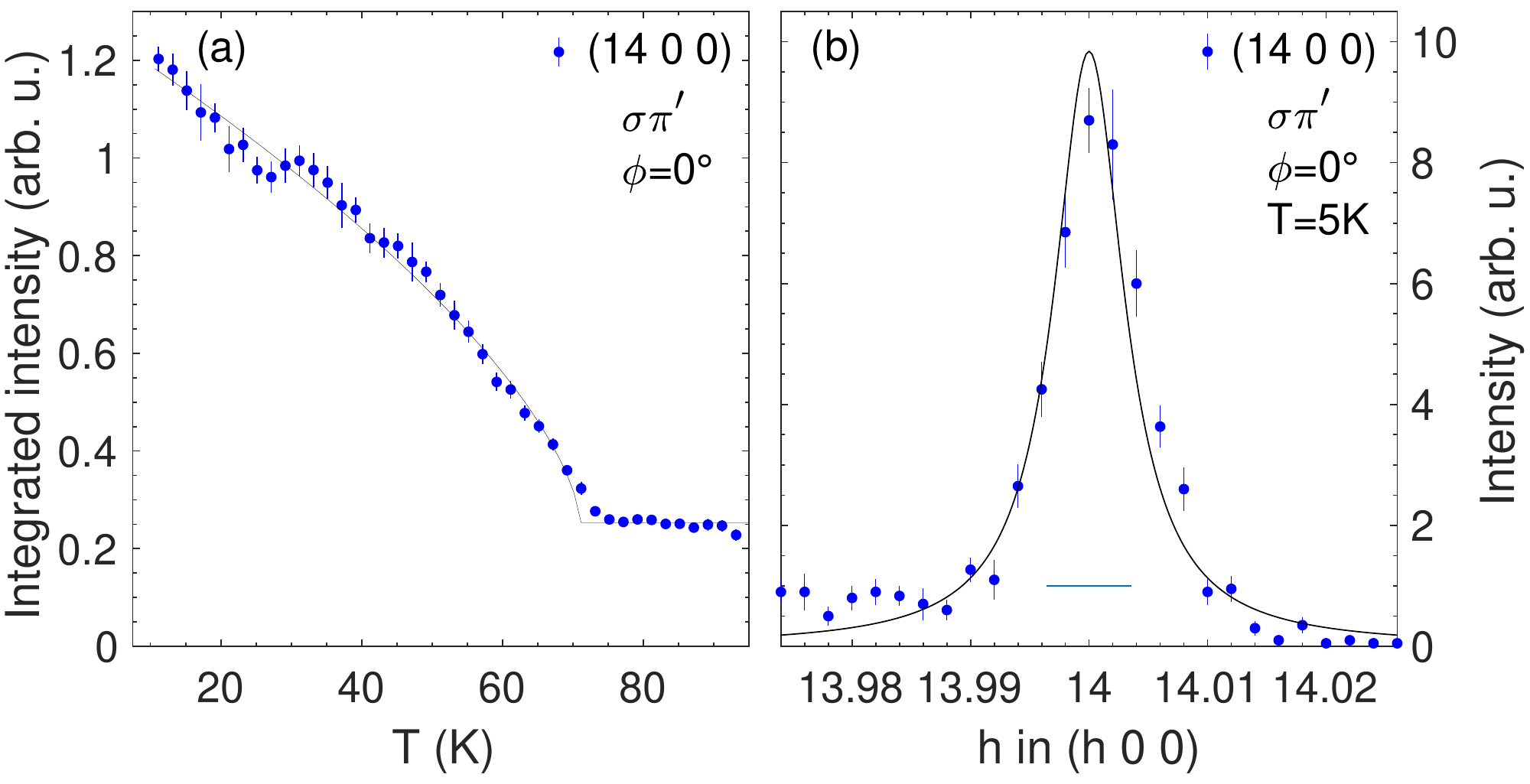}
\includegraphics[width=\linewidth]{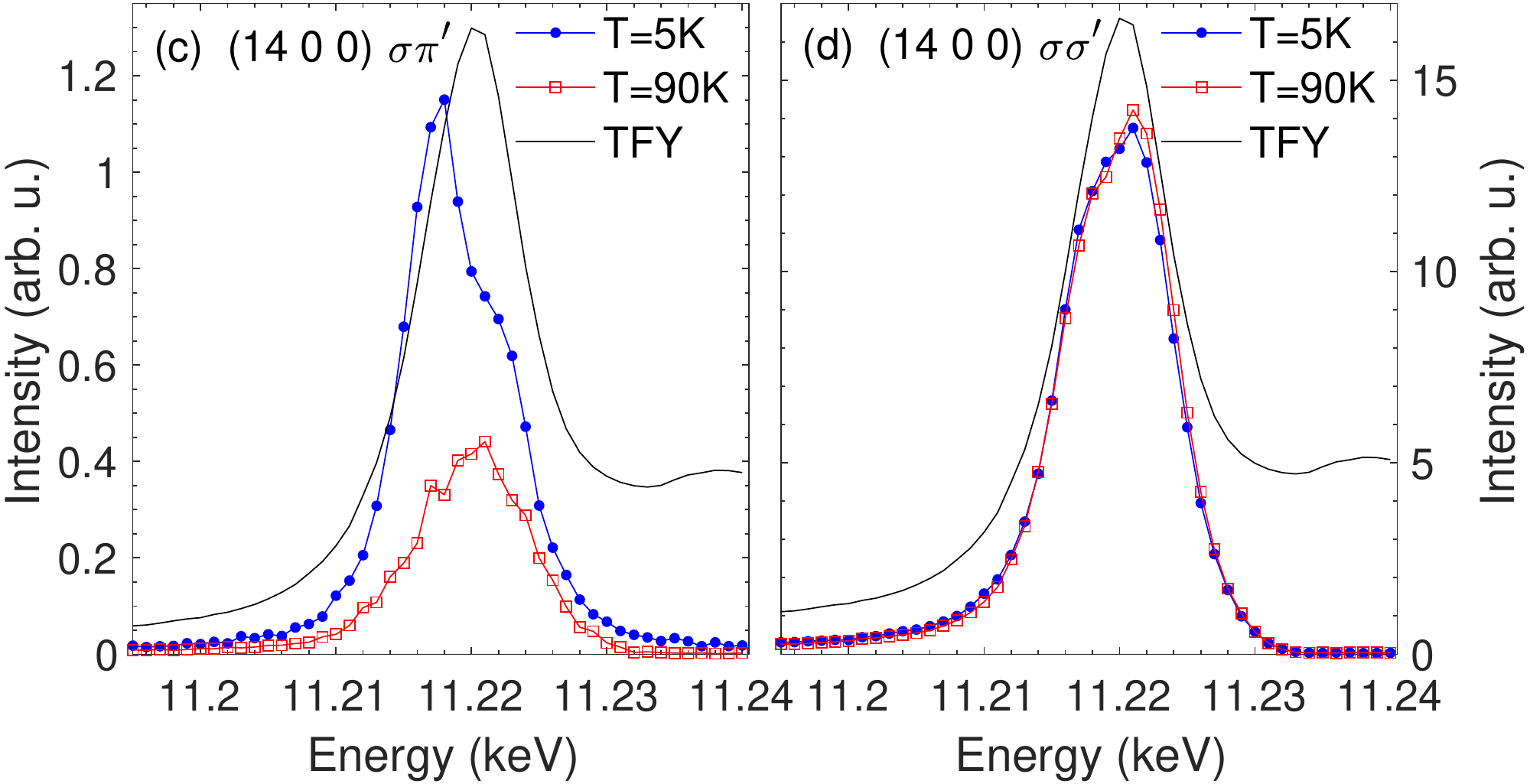}
\caption{REXS data of Tb$_{2+x}$Ir$_{2-x}$O$_{7-y}$ at the Ir $L_3$ edge. (a)~Temperature dependence of the (14 0 0) reflection in $\sigma\pi^{\prime}$ polarisation (blue circles), at an incident energy of $11.215\,$keV. The solid black line is a power-law fit to the data, yielding $T_{N} \sim 71$\,K. (b)~Reciprocal lattice scan along the $h$ direction of the (14 0 0) reflection (blue circles), and Lorentzian fit to the data (black solid line). (c, d)~Energy dependence of (14 0 0) reflection in $\sigma\pi^{\prime}$, $\sigma\sigma^{\prime}$ polarisations at $5\,$K (blue circles) and $90\,$K (red squares). The data were corrected for self-absorption. The black solid line is the Ir $L_3$ edge X-ray absorption spectrum, recorded in total fluorescence yield (TFY) mode.}
\label{Tb227_Tdep}
\end{figure}

Figure \ref{Tb227_Tdep} summarises key temperature, reciprocal space and energy dependences of the charge-forbidden (14 0 0) reflection at the Ir $L_3$ edge.
When cooling to $5\,$K, we observed a strong increase in intensity of the (14 0 0) reflection in $\sigma\pi^{\prime}$ polarisation, which suggests the formation of ${\bf k} = {\bf 0}$ magnetic order.
Although $(h00)$ reflections with $h = 4n + 2$ are forbidden for Thomson scattering, due to the presence of the diamond-glide plane symmetry of the pyrochlore lattice, on resonance one expects to observe anisotropic tensor susceptibility (ATS) scattering~\cite{dmitrienko2005polarization}.
As described in Ref.~\cite{donnerer2016all-in}, here the contribution from ATS scattering in the magnetically-relevant $\sigma\pi^{\prime}$ polarisation channel was suppressed by aligning the $(011)$ direction parallel to the scattering plane.
In this configuration, the magnetic signal of the (14 0 0) reflection can be isolated, which then exhibits a power-law like behaviour with temperature, with a N\'eel temperature of $\sim71\,$K (the data were corrected for beam-heating of $\sim 7\,$K), confirming the magnetic origin of the scattering.
The remaining intensity above the ordering temperature can be attributed to imperfect linear polarisation analysis of the scattered beam, which allows a fraction of the strong ATS signal in $\sigma\sigma^{\prime}$ polarisation to appear in the nominal $\sigma\pi^{\prime}$ polarisation channel (here referred to as ``leakage").

Reciprocal lattice scans confirm that the magnetic order is long-range, with a correlation length of at least a few hundred \AA{} [Fig.~\ref{Tb227_Tdep}(b)]. The magnetic origin of the signal is further supported by the energy dependence of the resonant signal. In the $\sigma\sigma^{\prime}$ channel, we find a temperature-independent two peak structure, usually attributed to REXS transitions to $t_{2g}$ and $e_g$ levels [Fig.~\ref{Tb227_Tdep}(d)]. Conversely, the energy profile in the $\sigma\pi^{\prime}$ channel shows a strong, temperature-dependent enhancement at the incident energy corresponding to transitions to $t_{2g}$ states (about $3\,$eV below the absorption maximum) [Fig.~\ref{Tb227_Tdep}(c)]. 
At high temperature, the energy profile resembles that of pure ATS scattering seen in $\sigma\sigma^{\prime}$ polarisation, suggesting that it originates from leakage of the strong ATS signal in $\sigma\sigma^{\prime}$ polarisation.
We conclude that below $\sim71$\,K, long-range ${\bf k} = {\bf 0}$ magnetic order of the Ir moments occurs in Tb$_{2+x}$Ir$_{2-x}$O$_{7-y}$.

\begin{figure}[htb]
\centering
\includegraphics[width=\linewidth]{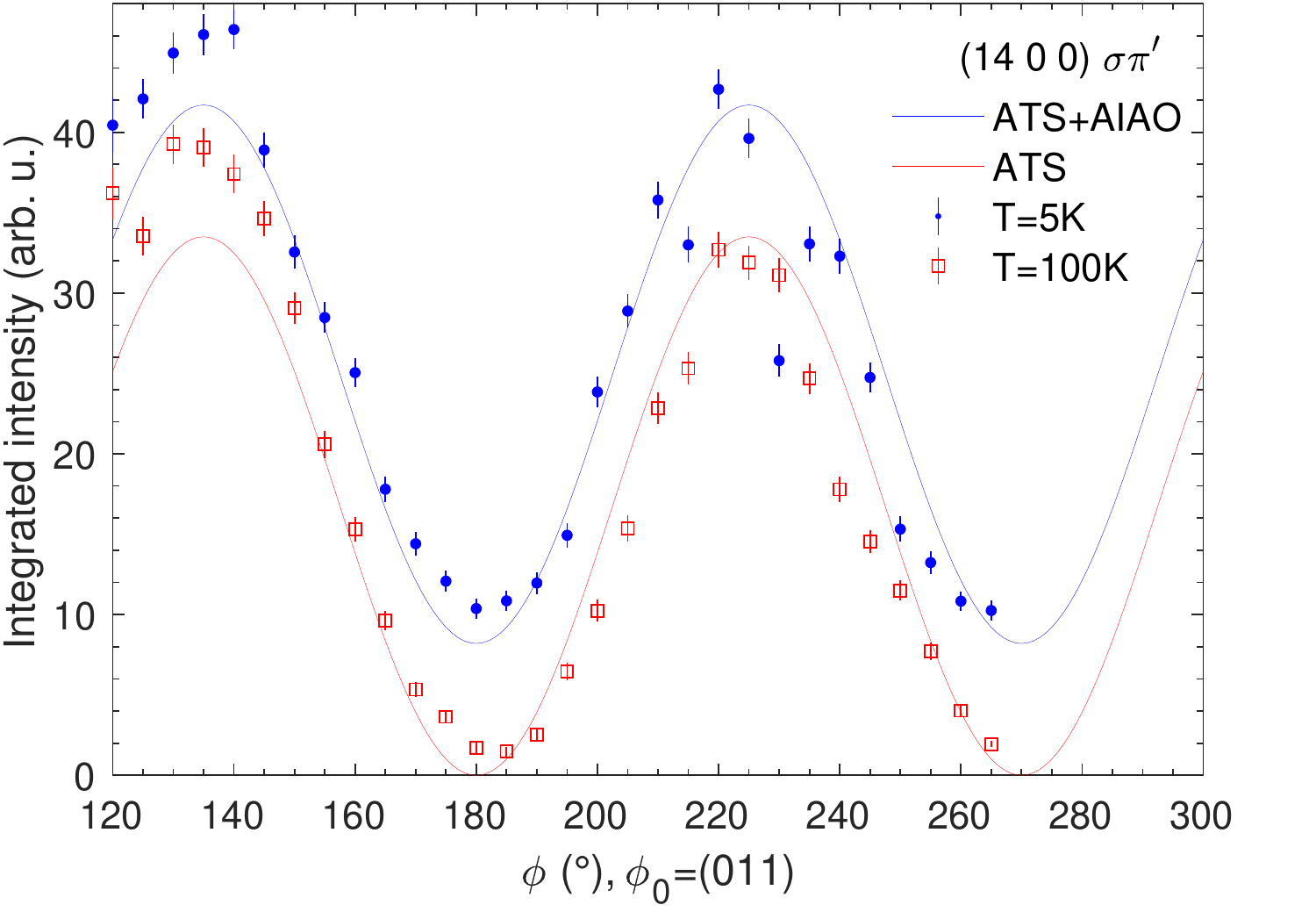}  
\caption{Integrated intensity of the (14 0 0) reflection  of Tb$_{2+x}$Ir$_{2-x}$O$_{7-y}$ as a function of azimuthal angle $\phi$ in $\sigma\pi^{\prime}$ polarisation at $5\,$K (blue circles) and $100\,$K (red squares), at an incident energy of $11.215\,$keV. The solid lines are calculated intensities for ATS scattering and resonant magnetic scattering from AIAO order (blue), and ATS scattering only (red).}
\label{Tb227_azi}
\end{figure}

Detecting a magnetic signal on forbidden $(h 0 0)$ reflections establishes a ${\bf k} = {\bf 0}$ propagation vector, which is compatible with different magnetic structures \cite{donnerer2016all-in}. 
To further distinguish candidate magnetic structures, we determined the orientation of the magnetic structure factor with an azimuthal scan. 
Figure \ref{Tb227_azi} shows the integrated intensities of rocking curves of the (14 0 0) reflection in $\sigma\pi^{\prime}$ polarisation, as a function of azimuthal angle $\phi$, at $5\,$K and $100\,$K. 
At both temperatures, the strongest contribution to the scattered intensity is a sinusoidal signal, originating from ATS scattering ($\propto \sin^2 2\phi$).
Additionally, we observe a temperature-dependent signal, which remains constant in $\phi$. 
The latter is the magnetic contribution, which shows that the magnetic structure factor lies parallel to the scattering vector ${\bf Q}$. This observation is only compatible with AIAO or $XY$ antiferromagnetic structures (see also Ref.~\cite{donnerer2016all-in}).
Given the NPD results of Refs.~\cite{lefrancois2015anisotropy, guo2017magnetic}, we consider the former to be more likely.

\subsubsection{Magnetic moment size}

Comparing the observed relative intensities of ATS and magnetic contributions to the ones expected from a calculation of the REXS cross-section offers the novel possibility of being able to estimate  the absolute size of the ordered Ir moment using resonant scattering; conventionally this can only be readily achieved using non-resonant techniques (see Appendix~\ref{moment_size_determination} and Ref.~\cite{donnerer2018magnetic}).
We here consider a minimal single-ion model of Ir$^{4+}$ in a cubic environment ($10Dq$), taking into account the spin-orbit coupling ($\zeta$) and trigonal fields ($\Delta$), and assuming that $\zeta, \Delta \ll 10Dq$~\cite{abragam1970electron, jackeli2009mott, ament2011theory, liu2012testing, ohgushi2013resonant, hozoi2014longer,moretti2014cairo}.
By comparison with RIXS experiments (see Refs.~\cite{hozoi2014longer, donnerer2018magnetic}), we can determine the parameters $\zeta$ and $\Delta$.
For Tb$_{2+x}$Ir$_{2-x}$O$_{7-y}$, we find that $\zeta$ and $\Delta$ are of roughly equal magnitude, with absolute values of $\sim450\,$meV.
In this case, the ground state wavefunction departs from the $J_\text{eff} = 1/2$ limit, which applies in the absence of trigonal distortion, and shows substantial hybridisation of $J_\text{eff} = 1/2$ and $J_\text{eff} = 3/2$ bands (see also Ref.~\cite{shinaoka2015phase}).
For this ground state, we find spin and orbital moments of $0.37\,\mu_B$ and $0.87\,\mu_B$, respectively, which yields a total magnetic moment of $\sim 1.61\,\mu_B$.

We then use this wavefunction to calculate REXS cross-sections at the Ir $L_3$ edge, assuming that 
$2p \rightarrow 5d$ dipole transitions are dominant.
This allows to determine the \emph{relative} contributions of resonant magnetic and ATS scattering to the intensity of the (14 0 0) reflection in $\sigma\pi^{\prime}$ polarisation. Noting that for $\sigma\pi^{\prime}$ polarisation, interference terms between magnetic and ATS signals cancel, we find a calculated intensity ratio of $I_{m}^{\sigma\pi^{\prime}} / I_{ats}^{\sigma\pi^{\prime}} \sim 5$, where $I_{m}^{\sigma\pi^{\prime}}$ and $I_{ats}^{\sigma\pi^{\prime}}$ are magnetic and ATS intensities, respectively.
From our REXS data, using the azimuthal scan of the (14 0 0) reflection at 5\,K (Fig. \ref{Tb227_azi}), by comparing intensities at peak and trough (corresponding to $I_{m}^{\sigma\pi^{\prime}} + I_{ats}^{\sigma\pi^{\prime}}$ and $I_{m}^{\sigma\pi^{\prime}}$, respectively), we find an experimental intensity ratio of $I_{m}^{\sigma\pi^{\prime}} / I_{ats}^{\sigma\pi^{\prime}} \sim 0.24$.
Given that $I_{m}^{\sigma\pi^{\prime}}$ is proportional to the square of the ordered moment, we propose that for Tb$_{2+x}$Ir$_{2-x}$O$_{7-y}$, the ordered magnetic moment is reduced to $\sim0.34\,\mu_B$ per Ir ion.

We identify two dominant sources of error in this estimate of the Ir magnetic moment size:
(1) The measured $I_{ats}^{\sigma\pi^{\prime}}$ is not only composed of a $5d(t_{2g})$ resonance ($11.215\,$keV), but also shows a contribution from a $5d(e_{g})$ resonance ($11.218\,$keV) and (2) the measured $I_m^{\sigma\pi^{\prime}}$ is not purely magnetic, but includes leakage from the strong ATS resonance in the $\sigma\sigma^{\prime}$ polarisation channel.
For (1), we estimated the contribution of the $5d(e_{g})$ resonance by fitting the $t_{2g}$ and $e_{g}$ resonances of the energy scan to Lorentzian functions. This indicates that $I_{ats}^{\sigma\pi^{\prime}}$ could be reduced by a factor of $\sim0.67$. 
For (2), we compared the magnetic intensity at low and high temperatures, which suggests that $I_m^{\sigma\pi^{\prime}}$ should be reduced by a factor of $\sim0.82$.
\footnote{The reason why we do not correct $I_{ats}$ for leakage is that (a) due to the stronger ATS signal this will comprise a much smaller relative error and (b) at the azimuthal angle where $I_{ats}^{\sigma\pi^{\prime}}$ is maximised, the corresponding signal in $I_{ats}^{\sigma\sigma^{\prime}}$ is minimised, which should substantially reduce the leakage.}
Applying these corrections to the experimental intensity ratio yields $I_{m}^{\sigma\pi^{\prime}} / I_{ats}^{\sigma\pi^{\prime}} \sim 0.29$, which translates to a magnetic moment of $\sim0.37\,\mu_B$ per Ir ion. While this is fairly imprecise, it gives an indication of the error associated with the estimated magnetic moment size.

Irrespective of the details of how the analysis is performed, our REXS experiments indicate that the magnitude of the Ir $5d$ moments in Tb$_{2+x}$Ir$_{2-x}$O$_{7+y}$ is strongly reduced relative to its value of $0.56\,\mu^{}_B$ in stoichiometric samples. This shows that stuffing can be an effective means to lower the size of the ordered moment. This result is of interest, as previous studies have suggested that a small ordered moment size, typically achieved by tuning the strength of electronic correlations, could be key to stabilising a WSM phase in pyrochlore iridates~\cite{go2012correlation, wang2017weyl}. However, while stuffing offers a means to suppress the magnetic order, it cannot directly be equated to tuning the electronic correlations, as it necessarily also introduces defects in the crystal structure. How this affects the predictions for novel topological states remains to be explored.

\subsection{REXS at the Tb $M_5$ edge}

\subsubsection{Magnetic structure}

\begin{figure}[htb]
\centering
\includegraphics[width=\linewidth]{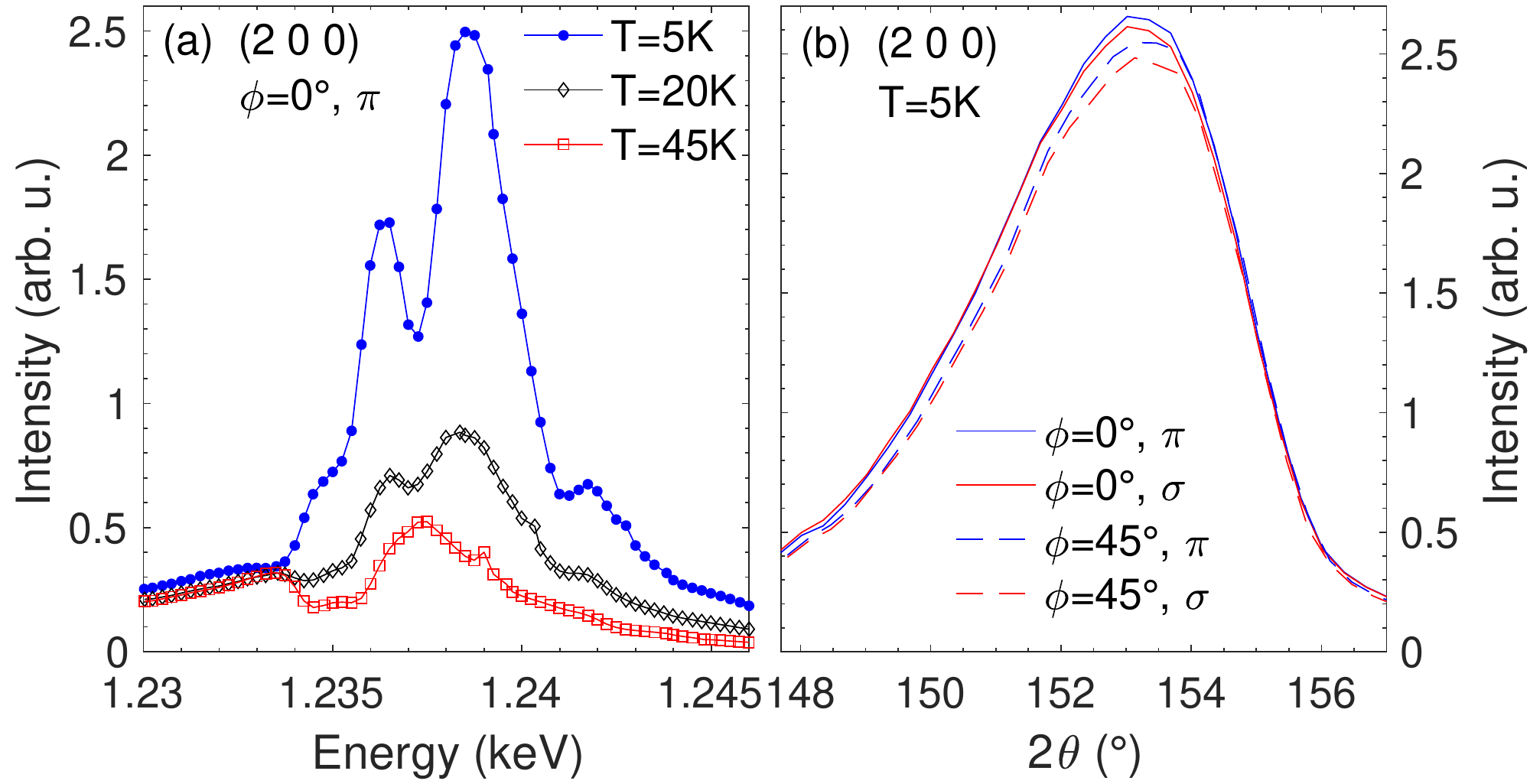}  
\caption{REXS data of Tb$_{2+x}$Ir$_{2-x}$O$_{7-y}$ at the Tb $M_5$ edge.
(a)~Energy scans as a function of temperature of the (2 0 0) reflection using $\pi$ incident polarisation. (b)~$\theta$--$2 \theta$ scans of the (2 0 0) reflection at $5\,$K, at azimuthal angles $\phi = 0^{\circ}$ (solid lines) and $\phi = 45^{\circ}$ (dashed lines), using $\pi$ (blue) and $\sigma$ (red) incident polarisations.}
\label{fig4}
\end{figure}

We now turn to the results of the REXS experiments performed in the vicinity of the Tb $M_5$ edge ($3d\rightarrow4f$ transitions). Figure~\ref{fig4}(a) shows photon energy scans for the charge-forbidden (2 0 0) reflection with $\pi$ incident polarisation, as a function of temperature. At 45\,K, a weak resonance is evident at $\sim1236\,$eV, associated with ATS processes. When cooling to 5\,K, we find a strong increase in intensity centred at $\sim1239\,$eV, arising from resonant magnetic scattering~\cite{goedkoop1988calculations}. This establishes unambiguously that the Tb sublattice also orders magnetically at low temperature, and with the same magnetic propagation vector of $\bf{k}=\bf{0}$ as that displayed by the Ir sublattice.

To further investigate the nature of this magnetic signal, we measured the intensity variation of the (2 0 0) magnetic peak as a function of the incident photon polarisation and sample azimuthal rotation. Figure~\ref{fig4}(b) displays the variation in intensity of the (2 0 0) magnetic reflection using $\sigma$ and $\pi$ incoming polarisation, at azimuthal angles of $\phi=0^{\circ}$ and $\phi=45^{\circ}$ [where $\phi=0^{\circ}$ corresponds to the (011) direction being parallel to the horizontal scattering plane]. We note that due to the cubic symmetry, covering an azimuthal range of 45 degrees should display all potential variation in scattered intensity.

As opposed to the REXS experiment at the Ir $L_3$ edge, in this experiment the outgoing polarisation was not analysed. In this case, the absence of any significant variation of the (2 0 0) peak with azimuth or incident photon polarization indicates that the Tb $4f$ magnetic structure factor vector lies parallel to the momentum transfer $\bf{Q}$ (see also Ref.~\cite{hill1996resonant}), as was found for the orientation of the Ir $5d$ magnetic structure factor. For the ATS contribution to the REXS signal, when no outgoing polarisation analysis is performed, for both $\sigma$ and $\pi$ incident polarisations, at a diffraction angle of $2 \theta \sim 152^{\circ}$ for the (2 0 0) reflection, the scattered intensity becomes virtually constant in azimuth (only $\sim10$ \% of the ATS intensity should show sinusoidal variation, with a period of $2 \phi$, see also Ref.~\cite{collins2001anisotropic}).

We also note that at the Tb $M_5$ edge, the magnetic signal is the dominant contribution to the resonant scattering (a factor of $\sim5$ stronger than the ATS scattering at 5\,K), opposite to what was observed at the Ir $L_3$ edge. Although the magnetic moment is expected to be substantially larger for Tb$^{3+}$ than for Ir$^{4+}$, this increase in scattered intensity is an electronic effect, which can intuitively be understood from the larger orbital overlap of initial and final states for $3d\rightarrow4f$ transitions, compared to $2p\rightarrow5d$ transitions at the Ir $L_3$ edge.

The most plausible explanation for these observations, taken in the round, is that the AIAO magnetic structure, displayed by both the Ir $5d$ and Tb $4f$ moments in Tb$_{2}$Ir$_{2}$O$_{7}$, is robust against significant levels ($x\sim0.18$) of stuffing Tb onto the Ir sites. 

\subsubsection{Magnetic structure in applied magnetic field}

\begin{figure}[htb]
\centering
\includegraphics[width=80mm]{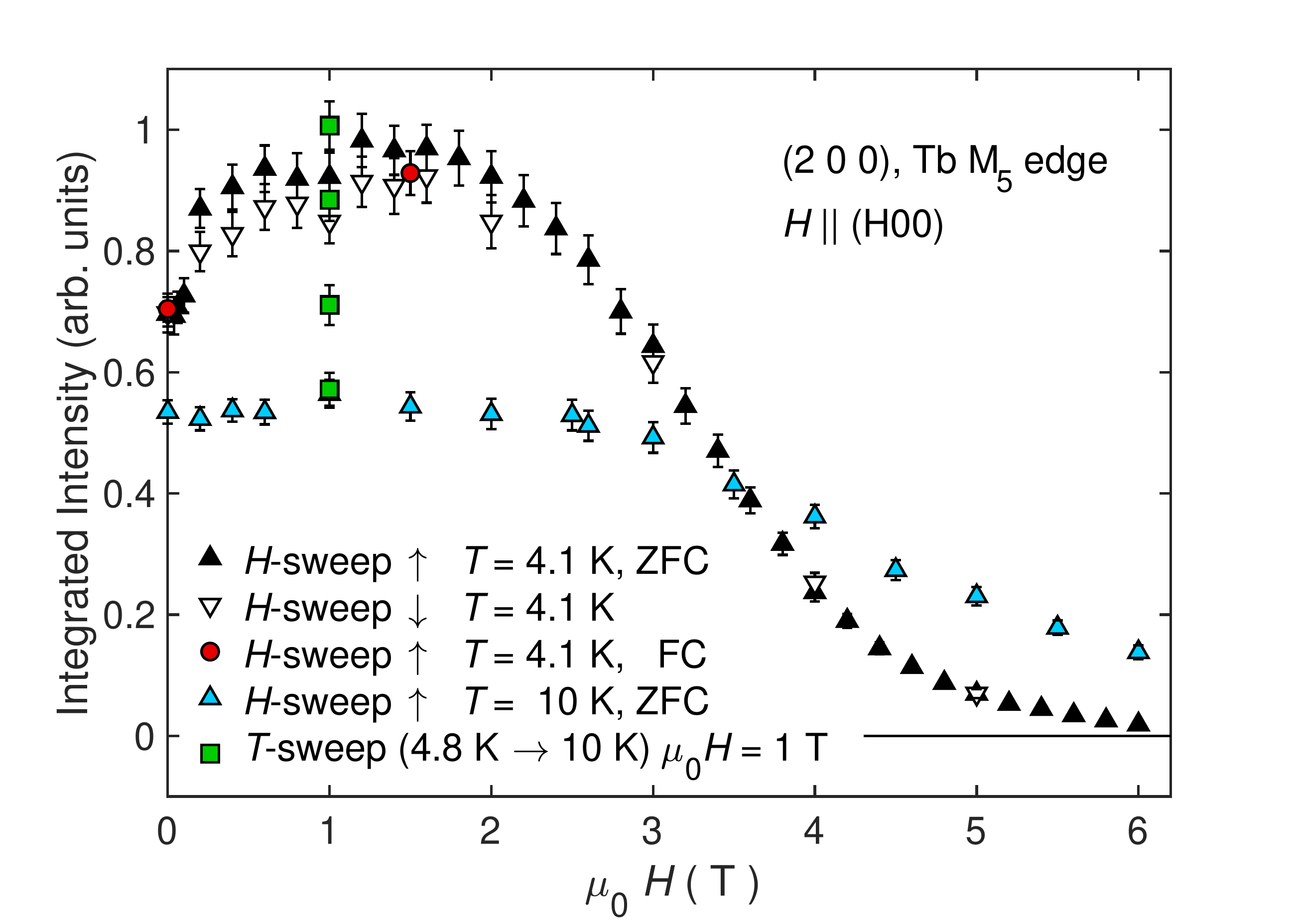}  
\caption{Integrated intensity of (200) REXS Bragg in magnetic fields of up to 6\,T applied along the (100) direction (Tb $M_5$ edge, $\pi$-incident polarization). The field-independent contribution due to ATS scattering has been subtracted. At 4.1\,K, the zero-field intensity increases by 30\% and forms a broad maximum around 1.5\,T (black triangles). There is no hysteresis (white triangles) or dependence on the cooling history (FC/ZFC, red circles). The non-monotonic behaviour appears to be suppressed at 10\,K.}
\label{fig5}
\end{figure}

We now discuss REXS experiments performed in applied magnetic fields. Figure~\ref{fig5} shows the field-dependent intensity of the (200) Tb magnetic Bragg peak, in fields of up to 6\,T, applied along the (100) direction. At 4.1\,K, the intensity increases relatively sharply at low fields, forming a broad maximum around $H_c\sim1.5$\,T, before it is suppressed at higher fields. Furthermore, this non-monotonic behaviour under applied field is not hysteretic or dependent on the cooling history, and is not observed at 10\,K.

We first compare our observations to magnetometry measurements of Tb$_{2}$Ir$_{2}$O$_{7}$ in the literature \cite{lefrancois2015anisotropy}. Below 10\,K, the magnetization exhibits an inflection point at $\sim 1.8$\,T, in close proximity to the intensity maximum we observe with REXS. This suggests that this behaviour is not unique to stuffed Tb$_{2+x}$Ir$_{2-x}$O$_{7-y}$, but also occurs in stoichiometric samples. The magnetization data show a tendency towards a saturation of the Tb moment at $5$\,T. This compares well to our observation that the (200) reflection becomes extinct at $6$\,T, suggesting that the Tb magnetic moments are fully polarised at this point.

The fact that the non-monotonic field-dependence of the REXS intensity could only be observed below 10\,K suggests an additional energy scale associated with Tb ions, such as $f$--$f$ interactions or single-ion effects. It is interesting to compare this result with NPD data of Ref.~\cite{guo2017magnetic}, where below 10\,K an additional (111) reflection appears, which should be absent in a pure realisation of AIAO magnetic order. The authors argued that at this temperature the magnetic structure slightly deviates from AIAO order, and acquires an additional $XY$ component. The proposed structure would correspond to the moments being tilted 6$^\circ$ away from the AIAO structure, in a fashion that leaves no net ferromagnetic moment in the unit cell. Such distortions and the observation of the (111) peak in NPD have a precedent in Tb$_2$Sn$_2$O$_7$, where Tb$^{3+}$ moments tilt 13$^{\circ}$ away from the 2I2O structure, albeit in a ferromagnetic fashion~\cite{mirebeau2005ordered}. 

We note that our zero-field REXS data does not allow to quantify such subtle modifications to the overall AIAO structure. However, a direct field-polarisation of any such magnetic order should still lead to a monotonic decrease of the intensity of the (200) reflection. We hence conjecture that a small applied field could suppress an $XY$ antiferromagnetic distortion, which would increase the intensity of the (200) reflection and restore pristine AIAO order at $H_c\sim1.5$\,T, before the magnetic moments become polarised at higher fields, and the (200) reflection disappears. The microscopic origin of such a behaviour is unclear, but potentially emerges out of a competition between $f$--$f$ exchange terms, the single-ion anisotropy, and the Zeeman coupling to the magnetic field. It is also known that Tb$^{3+}$ in pyrochlore compounds has a close-to-degenerate electric ground state \cite{aleksandrov1986crystal}, which could lead to a field-induced mixing of ground and first excited doublets that might alter the magnetic Hamiltonian.

\subsection{Temperature dependence of sublattice magnetisations}

\begin{figure}[htb]
\centering
\includegraphics[width=\linewidth]{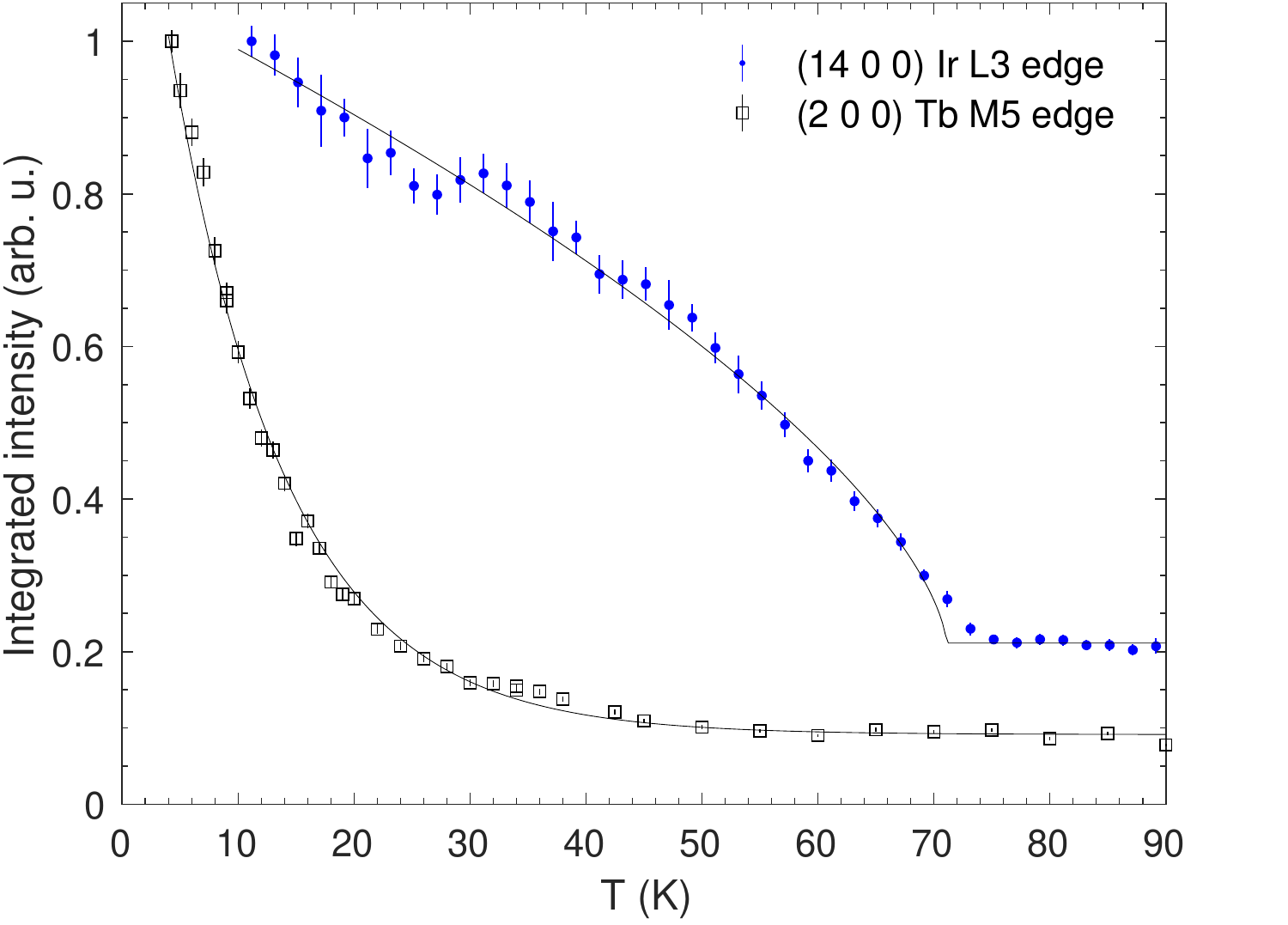}  
\caption{Integrated intensity of the (14 0 0) reflection at the Ir $L_3$ edge [blue dots, as shown in Fig.~\ref{Tb227_Tdep}(a)], and the (2 0 0) reflection at the Tb $M_5$ edge (black squares) of Tb$_{2+x}$Ir$_{2-x}$O$_{7-y}$, as a function of temperature.}
\label{Tdep-Tb-Ir}
\end{figure}

Finally, we consider the temperature development of magnetic order on the Tb and Ir sites in Tb$_{2+x}$Ir$_{2-x}$O$_{7-y}$. In Fig.~\ref{Tdep-Tb-Ir} we compare the temperature dependence of the magnetic Bragg peak intensities (proportional to the square of the staggered magnetisation) recorded at the Tb $M_5$ and Ir $L_3$ edges. While the temperature dependence of the Ir $5d$ staggered magnetic moment exhibits the typical order-parameter-like behaviour associated with spontaneous symmetry breaking below a thermally driven phase transition, the gradual development of the Tb $4f$ magnetic moment is more redolent of induced order, due to the presence of a mean field generated by the Ir moments. Thus our element selective REXS experiments reinforce the inferences made from NPD~\cite{lefrancois2015anisotropy, guo2017magnetic}, namely that the magnetic order in Tb$_{2}$Ir$_{2}$O$_{7}$ is primarily driven by the order on the Ir sublattice, hence establishing the primacy of $d$--$d$ interactions, with the Tb site ordering being induced by significant $d$--$f$ interactions.

\section{Conclusion}

In this study, we have used the element and electron shell specificity of REXS to selectively investigate the magnetic structures of the Tb $4f$ and Ir $5d$ sublattices in the stuffed pyrochlore iridate Tb$_{2+x}$Ir$_{2-x}$O$_{7-y}$. Our results establish that the stuffing of Tb ions onto Ir sites leads to significant reductions in both the N\'eel temperature and the staggered moment on the Ir sublattice. Most importantly, we have shown that the AIAO magnetic structure of the stoichiometric compound is robust against stuffing, at least up to a value of $x\sim0.18$. The significance of this result lies in the fact that stuffed pyrochlore iridates can thus retain the necessary requirements for a Weyl semimetal in this class of materials (overall cubic symmetry, in the presence of broken time-reversal symmetry from magnetic order), while at the same time the stuffing acts to suppress magnetic order, and hence the degree of electron correlations. We therefore propose that systematic studies of stuffed pyrochlore iridates may provide a viable route forward in the search for correlated materials with non-trivial band topology.

Data presented in this paper can be obtained from Ref.~\cite{donnerer2018dataset}

\begin{acknowledgments}

This work is supported by the UK Engineering and Physical Sciences Research Council under Grants No. EP/J016713/1 and No. EP/N027671/1.

\end{acknowledgments}

\appendix

\section{Magnetic moment size determination at the Ir $L_3$ edge of pyrochlore iridates}
\label{moment_size_determination}

We here argue that it is feasible to determine the magnetic moment size using REXS, by comparing the relative intensities of ATS and magnetic contributions to the ones expected from a model calculation. We consider a minimal single-ion model of Ir$^{4+}$ in a cubic environment ($10Dq$), taking into account the spin-orbit coupling ($\zeta$) and trigonal fields ($\Delta$, where a negative value corresponds to trigonal compression), as described in Refs.~\cite{abragam1970electron, jackeli2009mott, ament2011theory, liu2012testing, ohgushi2013resonant, hozoi2014longer, moretti2014cairo}.

As discussed in Ref.~\cite{donnerer2018magnetic}, we can use this model to derive the magnetic ($m$) and ATS ($ats$) contributions to REXS at the Ir $L_3$ edge, assuming that $2p \rightarrow 5d$ dipole transitions are dominant. For space group forbidden $(h00)$ reflections of pyrochlore iridates, where $h = 4n +2$, we find the following scattering amplitudes, written in the $2\times2$ Jones matrix formalism:

\[
\begin{aligned}
\mathcal{A}^{(h00)}_{ats} &= \frac{8(C_f^2+C_f-2)}{3(2+C_f^2)}
\begin{pmatrix}
\cos 2 \phi & -\sin \theta \sin 2 \phi  \\
\sin \theta \sin 2 \phi & \sin^2 \theta \cos 2 \phi \\
\end{pmatrix}\\
\mathcal{A}^{(h00)}_{m} &= \frac{4 [10 + (C_f-2) C_f]}{3 \sqrt{3} (2 + C_f^2)}
\begin{pmatrix}
0 & -i \sin \theta \\
-i \sin \theta & 0 
\end{pmatrix}
\end{aligned}
\]
where $2C_f=\delta-1+\sqrt{9+\delta(\delta-2)}$ and $\delta=2\Delta/\zeta$. $\theta$ is the Bragg angle, and we have defined the azimuthal angle $\phi=0$ when the crystallographic (011) direction is in the scattering plane.

A REXS experiment measures the coherent sum $I_{rexs}^{(h00)}~=~\left| \mathcal{A}^{(h00)}_{ats} +\mathcal{A}^{(h00)}_{m}  \right|^2$. For the $\sigma \pi^\prime$ polarisation channel, we find:

\begin{widetext}
\[
\begin{aligned}
I_{rexs}^{\sigma \pi^\prime} (h00) 
= &I_m^{\sigma \pi^\prime} (h00) &+& I_{ats}^{\sigma \pi^\prime} (h00) \\
= &\frac{16 [ 10+ (C_f-2)C_f]^2 \sin^2 \theta } {27(2+C_f^2)^2}
&+& \frac{64 ( C_f^2+C_f-2)^2 \sin^2 \theta \sin^2 2 \phi } {9 (2+C_f^2)^2}
\end{aligned}
\]
\end{widetext}
Note that interference terms between ATS and magnetic scattering cancel. By letting $\phi=\pi/4$, we derive the following intensity ratio of magnetic and ATS contributions:

\[
\label{calc_intensity_ratio}
\frac{I_m^{\sigma \pi} (h00)}{I_{ats}^{\sigma \pi} (h00)} = \frac{[10 + (C_f-2) C_f]^2}{12 (C_f^2 +C_f-2)^2}
\]

By comparing the measured intensity ratio ($meas$) to the one calculated ($calc$), we define a scaling factor $\alpha$:

\[
\left[ \frac{I_m^{\sigma \pi} (h00)}{I_{ats}^{\sigma \pi} (h00)} \right]_{meas} = 
\alpha \left[ \frac{I_m^{\sigma \pi} (h00)}{I_{ats}^{\sigma \pi} (h00)} \right]_{calc}
\]

We note that $I_m$ is proportional to the square of the single-ion magnetic moment, $\braket{\mu_z}^2$, where

\[
\braket{\mu_z}_{calc} = \braket{L_z} + 2\braket{S_z} = \frac{4 - C_f^2}{2 + C_f^2} \, \mu_B
\]
 
Thus we propose that we can use the scaling factor $\alpha$ to infer the measured magnetic moment size:

\[
\braket{\mu_z}_{meas} = \sqrt{ \alpha \braket{\mu_z}^2_{calc}}
\]

\subsection{Parameters used for Tb$_{2+x}$Ir$_{2-x}$O$_{7-y}$}

For Tb$_{2+x}$Ir$_{2-x}$O$_{7-y}$, the crystal-field excitations observed in RIXS experiments of Ref.~\cite{donnerer2018magnetic} allow to extract the single-ion parameters $\zeta = 449\,$meV and $\Delta = -469\,$meV. This yields $C_f  = 0.5494$ and $\braket{\mu_z}_{calc} \sim 1.61\,\mu_B$.


\begin{thebibliography}{49}%
\makeatletter
\providecommand \@ifxundefined [1]{%
 \@ifx{#1\undefined}
}%
\providecommand \@ifnum [1]{%
 \ifnum #1\expandafter \@firstoftwo
 \else \expandafter \@secondoftwo
 \fi
}%
\providecommand \@ifx [1]{%
 \ifx #1\expandafter \@firstoftwo
 \else \expandafter \@secondoftwo
 \fi
}%
\providecommand \natexlab [1]{#1}%
\providecommand \enquote  [1]{``#1''}%
\providecommand \bibnamefont  [1]{#1}%
\providecommand \bibfnamefont [1]{#1}%
\providecommand \citenamefont [1]{#1}%
\providecommand \href@noop [0]{\@secondoftwo}%
\providecommand \href [0]{\begingroup \@sanitize@url \@href}%
\providecommand \@href[1]{\@@startlink{#1}\@@href}%
\providecommand \@@href[1]{\endgroup#1\@@endlink}%
\providecommand \@sanitize@url [0]{\catcode `\\12\catcode `\$12\catcode
  `\&12\catcode `\#12\catcode `\^12\catcode `\_12\catcode `\%12\relax}%
\providecommand \@@startlink[1]{}%
\providecommand \@@endlink[0]{}%
\providecommand \url  [0]{\begingroup\@sanitize@url \@url }%
\providecommand \@url [1]{\endgroup\@href {#1}{\urlprefix }}%
\providecommand \urlprefix  [0]{URL }%
\providecommand \Eprint [0]{\href }%
\providecommand \doibase [0]{http://dx.doi.org/}%
\providecommand \selectlanguage [0]{\@gobble}%
\providecommand \bibinfo  [0]{\@secondoftwo}%
\providecommand \bibfield  [0]{\@secondoftwo}%
\providecommand \translation [1]{[#1]}%
\providecommand \BibitemOpen [0]{}%
\providecommand \bibitemStop [0]{}%
\providecommand \bibitemNoStop [0]{.\EOS\space}%
\providecommand \EOS [0]{\spacefactor3000\relax}%
\providecommand \BibitemShut  [1]{\csname bibitem#1\endcsname}%
\let\auto@bib@innerbib\@empty
\bibitem [{\citenamefont {Witczak-Krempa}\ \emph {et~al.}(2014)\citenamefont
  {Witczak-Krempa}, \citenamefont {Chen}, \citenamefont {Kim},\ and\
  \citenamefont {Balents}}]{krempa2014correlated}%
  \BibitemOpen
  \bibfield  {author} {\bibinfo {author} {\bibfnamefont {W.}~\bibnamefont
  {Witczak-Krempa}}, \bibinfo {author} {\bibfnamefont {G.}~\bibnamefont
  {Chen}}, \bibinfo {author} {\bibfnamefont {Y.~B.}\ \bibnamefont {Kim}}, \
  and\ \bibinfo {author} {\bibfnamefont {L.}~\bibnamefont {Balents}},\ }\href
  {\doibase 10.1146/annurev-conmatphys-020911-125138} {\bibfield  {journal}
  {\bibinfo  {journal} {Annu. Rev. of Condens. Matter Phys.}\ }\textbf
  {\bibinfo {volume} {5}},\ \bibinfo {pages} {57} (\bibinfo {year}
  {2014})}\BibitemShut {NoStop}%
\bibitem [{\citenamefont {Schaffer}\ \emph {et~al.}(2016)\citenamefont
  {Schaffer}, \citenamefont {Lee}, \citenamefont {Yang},\ and\ \citenamefont
  {Kim}}]{schaffer2016recent}%
  \BibitemOpen
  \bibfield  {author} {\bibinfo {author} {\bibfnamefont {R.}~\bibnamefont
  {Schaffer}}, \bibinfo {author} {\bibfnamefont {E.~K.-H.}\ \bibnamefont
  {Lee}}, \bibinfo {author} {\bibfnamefont {B.-J.}\ \bibnamefont {Yang}}, \
  and\ \bibinfo {author} {\bibfnamefont {Y.~B.}\ \bibnamefont {Kim}},\ }\href
  {http://stacks.iop.org/0034-4885/79/i=9/a=094504} {\bibfield  {journal}
  {\bibinfo  {journal} {Rep. Prog. Phys.}\ }\textbf {\bibinfo {volume} {79}},\
  \bibinfo {pages} {094504} (\bibinfo {year} {2016})}\BibitemShut {NoStop}%
\bibitem [{\citenamefont {Wan}\ \emph {et~al.}(2011)\citenamefont {Wan},
  \citenamefont {Turner}, \citenamefont {Vishwanath},\ and\ \citenamefont
  {Savrasov}}]{wan2011topological}%
  \BibitemOpen
  \bibfield  {author} {\bibinfo {author} {\bibfnamefont {X.}~\bibnamefont
  {Wan}}, \bibinfo {author} {\bibfnamefont {A.~M.}\ \bibnamefont {Turner}},
  \bibinfo {author} {\bibfnamefont {A.}~\bibnamefont {Vishwanath}}, \ and\
  \bibinfo {author} {\bibfnamefont {S.~Y.}\ \bibnamefont {Savrasov}},\ }\href
  {\doibase 10.1103/PhysRevB.83.205101} {\bibfield  {journal} {\bibinfo
  {journal} {Phys. Rev. B}\ }\textbf {\bibinfo {volume} {83}},\ \bibinfo
  {pages} {205101} (\bibinfo {year} {2011})}\BibitemShut {NoStop}%
\bibitem [{\citenamefont {Witczak-Krempa}\ and\ \citenamefont
  {Kim}(2012)}]{krempa2012topological}%
  \BibitemOpen
  \bibfield  {author} {\bibinfo {author} {\bibfnamefont {W.}~\bibnamefont
  {Witczak-Krempa}}\ and\ \bibinfo {author} {\bibfnamefont {Y.~B.}\
  \bibnamefont {Kim}},\ }\href {\doibase 10.1103/PhysRevB.85.045124} {\bibfield
   {journal} {\bibinfo  {journal} {Phys. Rev. B}\ }\textbf {\bibinfo {volume}
  {85}},\ \bibinfo {pages} {045124} (\bibinfo {year} {2012})}\BibitemShut
  {NoStop}%
\bibitem [{\citenamefont {Go}\ \emph {et~al.}(2012)\citenamefont {Go},
  \citenamefont {Witczak-Krempa}, \citenamefont {Jeon}, \citenamefont {Park},\
  and\ \citenamefont {Kim}}]{go2012correlation}%
  \BibitemOpen
  \bibfield  {author} {\bibinfo {author} {\bibfnamefont {A.}~\bibnamefont
  {Go}}, \bibinfo {author} {\bibfnamefont {W.}~\bibnamefont {Witczak-Krempa}},
  \bibinfo {author} {\bibfnamefont {G.~S.}\ \bibnamefont {Jeon}}, \bibinfo
  {author} {\bibfnamefont {K.}~\bibnamefont {Park}}, \ and\ \bibinfo {author}
  {\bibfnamefont {Y.~B.}\ \bibnamefont {Kim}},\ }\href {\doibase
  10.1103/PhysRevLett.109.066401} {\bibfield  {journal} {\bibinfo  {journal}
  {Phys. Rev. Lett.}\ }\textbf {\bibinfo {volume} {109}},\ \bibinfo {pages}
  {066401} (\bibinfo {year} {2012})}\BibitemShut {NoStop}%
\bibitem [{\citenamefont {Witczak-Krempa}\ \emph {et~al.}(2013)\citenamefont
  {Witczak-Krempa}, \citenamefont {Go},\ and\ \citenamefont
  {Kim}}]{krempa2013pyrochlore}%
  \BibitemOpen
  \bibfield  {author} {\bibinfo {author} {\bibfnamefont {W.}~\bibnamefont
  {Witczak-Krempa}}, \bibinfo {author} {\bibfnamefont {A.}~\bibnamefont {Go}},
  \ and\ \bibinfo {author} {\bibfnamefont {Y.~B.}\ \bibnamefont {Kim}},\ }\href
  {\doibase 10.1103/PhysRevB.87.155101} {\bibfield  {journal} {\bibinfo
  {journal} {Phys. Rev. B}\ }\textbf {\bibinfo {volume} {87}},\ \bibinfo
  {pages} {155101} (\bibinfo {year} {2013})}\BibitemShut {NoStop}%
\bibitem [{\citenamefont {Lv}\ \emph {et~al.}(2015{\natexlab{a}})\citenamefont
  {Lv}, \citenamefont {Weng}, \citenamefont {Fu}, \citenamefont {Wang},
  \citenamefont {Miao}, \citenamefont {Ma}, \citenamefont {Richard},
  \citenamefont {Huang}, \citenamefont {Zhao}, \citenamefont {Chen},
  \citenamefont {Fang}, \citenamefont {Dai}, \citenamefont {Qian},\ and\
  \citenamefont {Ding}}]{lv2015experimental}%
  \BibitemOpen
  \bibfield  {author} {\bibinfo {author} {\bibfnamefont {B.~Q.}\ \bibnamefont
  {Lv}}, \bibinfo {author} {\bibfnamefont {H.~M.}\ \bibnamefont {Weng}},
  \bibinfo {author} {\bibfnamefont {B.~B.}\ \bibnamefont {Fu}}, \bibinfo
  {author} {\bibfnamefont {X.~P.}\ \bibnamefont {Wang}}, \bibinfo {author}
  {\bibfnamefont {H.}~\bibnamefont {Miao}}, \bibinfo {author} {\bibfnamefont
  {J.}~\bibnamefont {Ma}}, \bibinfo {author} {\bibfnamefont {P.}~\bibnamefont
  {Richard}}, \bibinfo {author} {\bibfnamefont {X.~C.}\ \bibnamefont {Huang}},
  \bibinfo {author} {\bibfnamefont {L.~X.}\ \bibnamefont {Zhao}}, \bibinfo
  {author} {\bibfnamefont {G.~F.}\ \bibnamefont {Chen}}, \bibinfo {author}
  {\bibfnamefont {Z.}~\bibnamefont {Fang}}, \bibinfo {author} {\bibfnamefont
  {X.}~\bibnamefont {Dai}}, \bibinfo {author} {\bibfnamefont {T.}~\bibnamefont
  {Qian}}, \ and\ \bibinfo {author} {\bibfnamefont {H.}~\bibnamefont {Ding}},\
  }\href {\doibase 10.1103/PhysRevX.5.031013} {\bibfield  {journal} {\bibinfo
  {journal} {Phys. Rev. X}\ }\textbf {\bibinfo {volume} {5}},\ \bibinfo {pages}
  {031013} (\bibinfo {year} {2015}{\natexlab{a}})}\BibitemShut {NoStop}%
\bibitem [{\citenamefont {Lv}\ \emph {et~al.}(2015{\natexlab{b}})\citenamefont
  {Lv}, \citenamefont {Xu}, \citenamefont {Weng}, \citenamefont {Ma},
  \citenamefont {Richard}, \citenamefont {Huang}, \citenamefont {Zhao},
  \citenamefont {Chen}, \citenamefont {Matt}, \citenamefont {Bisti},
  \citenamefont {Strocov}, \citenamefont {Mesot}, \citenamefont {Fang},
  \citenamefont {Dai}, \citenamefont {Qian}, \citenamefont {Shi},\ and\
  \citenamefont {Ding}}]{lv2015observation}%
  \BibitemOpen
  \bibfield  {author} {\bibinfo {author} {\bibfnamefont {B.~Q.}\ \bibnamefont
  {Lv}}, \bibinfo {author} {\bibfnamefont {N.}~\bibnamefont {Xu}}, \bibinfo
  {author} {\bibfnamefont {H.~M.}\ \bibnamefont {Weng}}, \bibinfo {author}
  {\bibfnamefont {J.~Z.}\ \bibnamefont {Ma}}, \bibinfo {author} {\bibfnamefont
  {P.}~\bibnamefont {Richard}}, \bibinfo {author} {\bibfnamefont {X.~C.}\
  \bibnamefont {Huang}}, \bibinfo {author} {\bibfnamefont {L.~X.}\ \bibnamefont
  {Zhao}}, \bibinfo {author} {\bibfnamefont {G.~F.}\ \bibnamefont {Chen}},
  \bibinfo {author} {\bibfnamefont {C.~E.}\ \bibnamefont {Matt}}, \bibinfo
  {author} {\bibfnamefont {F.}~\bibnamefont {Bisti}}, \bibinfo {author}
  {\bibfnamefont {V.~N.}\ \bibnamefont {Strocov}}, \bibinfo {author}
  {\bibfnamefont {J.}~\bibnamefont {Mesot}}, \bibinfo {author} {\bibfnamefont
  {Z.}~\bibnamefont {Fang}}, \bibinfo {author} {\bibfnamefont {X.}~\bibnamefont
  {Dai}}, \bibinfo {author} {\bibfnamefont {T.}~\bibnamefont {Qian}}, \bibinfo
  {author} {\bibfnamefont {M.}~\bibnamefont {Shi}}, \ and\ \bibinfo {author}
  {\bibfnamefont {H.}~\bibnamefont {Ding}},\ }\href
  {http://dx.doi.org/10.1038/nphys3426} {\bibfield  {journal} {\bibinfo
  {journal} {Nat. Phys.}\ }\textbf {\bibinfo {volume} {11}},\ \bibinfo {pages}
  {724} (\bibinfo {year} {2015}{\natexlab{b}})}\BibitemShut {NoStop}%
\bibitem [{\citenamefont {Xu}\ \emph {et~al.}(2015{\natexlab{a}})\citenamefont
  {Xu}, \citenamefont {Belopolski}, \citenamefont {Alidoust}, \citenamefont
  {Neupane}, \citenamefont {Bian}, \citenamefont {Zhang}, \citenamefont
  {Sankar}, \citenamefont {Chang}, \citenamefont {Yuan}, \citenamefont {Lee},
  \citenamefont {Huang}, \citenamefont {Zheng}, \citenamefont {Ma},
  \citenamefont {Sanchez}, \citenamefont {Wang}, \citenamefont {Bansil},
  \citenamefont {Chou}, \citenamefont {Shibayev}, \citenamefont {Lin},
  \citenamefont {Jia},\ and\ \citenamefont {Hasan}}]{xu2015discovery1}%
  \BibitemOpen
  \bibfield  {author} {\bibinfo {author} {\bibfnamefont {S.-Y.}\ \bibnamefont
  {Xu}}, \bibinfo {author} {\bibfnamefont {I.}~\bibnamefont {Belopolski}},
  \bibinfo {author} {\bibfnamefont {N.}~\bibnamefont {Alidoust}}, \bibinfo
  {author} {\bibfnamefont {M.}~\bibnamefont {Neupane}}, \bibinfo {author}
  {\bibfnamefont {G.}~\bibnamefont {Bian}}, \bibinfo {author} {\bibfnamefont
  {C.}~\bibnamefont {Zhang}}, \bibinfo {author} {\bibfnamefont
  {R.}~\bibnamefont {Sankar}}, \bibinfo {author} {\bibfnamefont
  {G.}~\bibnamefont {Chang}}, \bibinfo {author} {\bibfnamefont
  {Z.}~\bibnamefont {Yuan}}, \bibinfo {author} {\bibfnamefont {C.-C.}\
  \bibnamefont {Lee}}, \bibinfo {author} {\bibfnamefont {S.-M.}\ \bibnamefont
  {Huang}}, \bibinfo {author} {\bibfnamefont {H.}~\bibnamefont {Zheng}},
  \bibinfo {author} {\bibfnamefont {J.}~\bibnamefont {Ma}}, \bibinfo {author}
  {\bibfnamefont {D.~S.}\ \bibnamefont {Sanchez}}, \bibinfo {author}
  {\bibfnamefont {B.}~\bibnamefont {Wang}}, \bibinfo {author} {\bibfnamefont
  {A.}~\bibnamefont {Bansil}}, \bibinfo {author} {\bibfnamefont
  {F.}~\bibnamefont {Chou}}, \bibinfo {author} {\bibfnamefont {P.~P.}\
  \bibnamefont {Shibayev}}, \bibinfo {author} {\bibfnamefont {H.}~\bibnamefont
  {Lin}}, \bibinfo {author} {\bibfnamefont {S.}~\bibnamefont {Jia}}, \ and\
  \bibinfo {author} {\bibfnamefont {M.~Z.}\ \bibnamefont {Hasan}},\ }\href
  {\doibase 10.1126/science.aaa9297} {\bibfield  {journal} {\bibinfo  {journal}
  {Science}\ }\textbf {\bibinfo {volume} {349}},\ \bibinfo {pages} {613}
  (\bibinfo {year} {2015}{\natexlab{a}})}\BibitemShut {NoStop}%
\bibitem [{\citenamefont {Xu}\ \emph {et~al.}(2015{\natexlab{b}})\citenamefont
  {Xu}, \citenamefont {Alidoust}, \citenamefont {Belopolski}, \citenamefont
  {Yuan}, \citenamefont {Bian}, \citenamefont {Chang}, \citenamefont {Zheng},
  \citenamefont {Strocov}, \citenamefont {Sanchez}, \citenamefont {Chang},
  \citenamefont {Zhang}, \citenamefont {Mou}, \citenamefont {Wu}, \citenamefont
  {Huang}, \citenamefont {Lee}, \citenamefont {Huang}, \citenamefont {Wang},
  \citenamefont {Bansil}, \citenamefont {Jeng}, \citenamefont {Neupert},
  \citenamefont {Kaminski}, \citenamefont {Lin}, \citenamefont {Jia},\ and\
  \citenamefont {Zahid~Hasan}}]{xu2015discovery2}%
  \BibitemOpen
  \bibfield  {author} {\bibinfo {author} {\bibfnamefont {S.-Y.}\ \bibnamefont
  {Xu}}, \bibinfo {author} {\bibfnamefont {N.}~\bibnamefont {Alidoust}},
  \bibinfo {author} {\bibfnamefont {I.}~\bibnamefont {Belopolski}}, \bibinfo
  {author} {\bibfnamefont {Z.}~\bibnamefont {Yuan}}, \bibinfo {author}
  {\bibfnamefont {G.}~\bibnamefont {Bian}}, \bibinfo {author} {\bibfnamefont
  {T.-R.}\ \bibnamefont {Chang}}, \bibinfo {author} {\bibfnamefont
  {H.}~\bibnamefont {Zheng}}, \bibinfo {author} {\bibfnamefont {V.~N.}\
  \bibnamefont {Strocov}}, \bibinfo {author} {\bibfnamefont {D.~S.}\
  \bibnamefont {Sanchez}}, \bibinfo {author} {\bibfnamefont {G.}~\bibnamefont
  {Chang}}, \bibinfo {author} {\bibfnamefont {C.}~\bibnamefont {Zhang}},
  \bibinfo {author} {\bibfnamefont {D.}~\bibnamefont {Mou}}, \bibinfo {author}
  {\bibfnamefont {Y.}~\bibnamefont {Wu}}, \bibinfo {author} {\bibfnamefont
  {L.}~\bibnamefont {Huang}}, \bibinfo {author} {\bibfnamefont {C.-C.}\
  \bibnamefont {Lee}}, \bibinfo {author} {\bibfnamefont {S.-M.}\ \bibnamefont
  {Huang}}, \bibinfo {author} {\bibfnamefont {B.}~\bibnamefont {Wang}},
  \bibinfo {author} {\bibfnamefont {A.}~\bibnamefont {Bansil}}, \bibinfo
  {author} {\bibfnamefont {H.-T.}\ \bibnamefont {Jeng}}, \bibinfo {author}
  {\bibfnamefont {T.}~\bibnamefont {Neupert}}, \bibinfo {author} {\bibfnamefont
  {A.}~\bibnamefont {Kaminski}}, \bibinfo {author} {\bibfnamefont
  {H.}~\bibnamefont {Lin}}, \bibinfo {author} {\bibfnamefont {S.}~\bibnamefont
  {Jia}}, \ and\ \bibinfo {author} {\bibfnamefont {M.}~\bibnamefont
  {Zahid~Hasan}},\ }\href {http://dx.doi.org/10.1038/nphys3437} {\bibfield
  {journal} {\bibinfo  {journal} {Nat. Phys.}\ }\textbf {\bibinfo {volume}
  {11}},\ \bibinfo {pages} {748} (\bibinfo {year}
  {2015}{\natexlab{b}})}\BibitemShut {NoStop}%
\bibitem [{\citenamefont {Yang}\ \emph {et~al.}(2015)\citenamefont {Yang},
  \citenamefont {Liu}, \citenamefont {Sun}, \citenamefont {Peng}, \citenamefont
  {Yang}, \citenamefont {Zhang}, \citenamefont {Zhou}, \citenamefont {Zhang},
  \citenamefont {Guo}, \citenamefont {Rahn}, \citenamefont {Prabhakaran},
  \citenamefont {Hussain}, \citenamefont {Mo}, \citenamefont {Felser},
  \citenamefont {Yan},\ and\ \citenamefont {Chen}}]{yang2015weyl}%
  \BibitemOpen
  \bibfield  {author} {\bibinfo {author} {\bibfnamefont {L.~X.}\ \bibnamefont
  {Yang}}, \bibinfo {author} {\bibfnamefont {Z.~K.}\ \bibnamefont {Liu}},
  \bibinfo {author} {\bibfnamefont {Y.}~\bibnamefont {Sun}}, \bibinfo {author}
  {\bibfnamefont {H.}~\bibnamefont {Peng}}, \bibinfo {author} {\bibfnamefont
  {H.~F.}\ \bibnamefont {Yang}}, \bibinfo {author} {\bibfnamefont
  {T.}~\bibnamefont {Zhang}}, \bibinfo {author} {\bibfnamefont
  {B.}~\bibnamefont {Zhou}}, \bibinfo {author} {\bibfnamefont {Y.}~\bibnamefont
  {Zhang}}, \bibinfo {author} {\bibfnamefont {Y.~F.}\ \bibnamefont {Guo}},
  \bibinfo {author} {\bibfnamefont {M.}~\bibnamefont {Rahn}}, \bibinfo {author}
  {\bibfnamefont {D.}~\bibnamefont {Prabhakaran}}, \bibinfo {author}
  {\bibfnamefont {Z.}~\bibnamefont {Hussain}}, \bibinfo {author} {\bibfnamefont
  {S.~K.}\ \bibnamefont {Mo}}, \bibinfo {author} {\bibfnamefont
  {C.}~\bibnamefont {Felser}}, \bibinfo {author} {\bibfnamefont
  {B.}~\bibnamefont {Yan}}, \ and\ \bibinfo {author} {\bibfnamefont {Y.~L.}\
  \bibnamefont {Chen}},\ }\href {http://dx.doi.org/10.1038/nphys3425}
  {\bibfield  {journal} {\bibinfo  {journal} {Nat. Phys.}\ }\textbf {\bibinfo
  {volume} {11}},\ \bibinfo {pages} {728} (\bibinfo {year} {2015})}\BibitemShut
  {NoStop}%
\bibitem [{\citenamefont {Shinaoka}\ \emph {et~al.}(2015)\citenamefont
  {Shinaoka}, \citenamefont {Hoshino}, \citenamefont {Troyer},\ and\
  \citenamefont {Werner}}]{shinaoka2015phase}%
  \BibitemOpen
  \bibfield  {author} {\bibinfo {author} {\bibfnamefont {H.}~\bibnamefont
  {Shinaoka}}, \bibinfo {author} {\bibfnamefont {S.}~\bibnamefont {Hoshino}},
  \bibinfo {author} {\bibfnamefont {M.}~\bibnamefont {Troyer}}, \ and\ \bibinfo
  {author} {\bibfnamefont {P.}~\bibnamefont {Werner}},\ }\href {\doibase
  10.1103/PhysRevLett.115.156401} {\bibfield  {journal} {\bibinfo  {journal}
  {Phys. Rev. Lett.}\ }\textbf {\bibinfo {volume} {115}},\ \bibinfo {pages}
  {156401} (\bibinfo {year} {2015})}\BibitemShut {NoStop}%
\bibitem [{\citenamefont {Zhang}\ \emph {et~al.}(2017)\citenamefont {Zhang},
  \citenamefont {Haule},\ and\ \citenamefont
  {Vanderbilt}}]{zhang2017metal-insulator}%
  \BibitemOpen
  \bibfield  {author} {\bibinfo {author} {\bibfnamefont {H.}~\bibnamefont
  {Zhang}}, \bibinfo {author} {\bibfnamefont {K.}~\bibnamefont {Haule}}, \ and\
  \bibinfo {author} {\bibfnamefont {D.}~\bibnamefont {Vanderbilt}},\ }\href
  {\doibase 10.1103/PhysRevLett.118.026404} {\bibfield  {journal} {\bibinfo
  {journal} {Phys. Rev. Lett.}\ }\textbf {\bibinfo {volume} {118}},\ \bibinfo
  {pages} {026404} (\bibinfo {year} {2017})}\BibitemShut {NoStop}%
\bibitem [{\citenamefont {Yang}\ and\ \citenamefont
  {Nagaosa}(2014)}]{yang2014emergent}%
  \BibitemOpen
  \bibfield  {author} {\bibinfo {author} {\bibfnamefont {B.-J.}\ \bibnamefont
  {Yang}}\ and\ \bibinfo {author} {\bibfnamefont {N.}~\bibnamefont {Nagaosa}},\
  }\href {\doibase 10.1103/PhysRevLett.112.246402} {\bibfield  {journal}
  {\bibinfo  {journal} {Phys. Rev. Lett.}\ }\textbf {\bibinfo {volume} {112}},\
  \bibinfo {pages} {246402} (\bibinfo {year} {2014})}\BibitemShut {NoStop}%
\bibitem [{\citenamefont {Zhao}\ \emph {et~al.}(2011)\citenamefont {Zhao},
  \citenamefont {Mackie}, \citenamefont {MacLaughlin}, \citenamefont {Bernal},
  \citenamefont {Ishikawa}, \citenamefont {Ohta},\ and\ \citenamefont
  {Nakatsuji}}]{zhao2011magnetic}%
  \BibitemOpen
  \bibfield  {author} {\bibinfo {author} {\bibfnamefont {S.}~\bibnamefont
  {Zhao}}, \bibinfo {author} {\bibfnamefont {J.~M.}\ \bibnamefont {Mackie}},
  \bibinfo {author} {\bibfnamefont {D.~E.}\ \bibnamefont {MacLaughlin}},
  \bibinfo {author} {\bibfnamefont {O.~O.}\ \bibnamefont {Bernal}}, \bibinfo
  {author} {\bibfnamefont {J.~J.}\ \bibnamefont {Ishikawa}}, \bibinfo {author}
  {\bibfnamefont {Y.}~\bibnamefont {Ohta}}, \ and\ \bibinfo {author}
  {\bibfnamefont {S.}~\bibnamefont {Nakatsuji}},\ }\href {\doibase
  10.1103/PhysRevB.83.180402} {\bibfield  {journal} {\bibinfo  {journal} {Phys.
  Rev. B}\ }\textbf {\bibinfo {volume} {83}},\ \bibinfo {pages} {180402}
  (\bibinfo {year} {2011})}\BibitemShut {NoStop}%
\bibitem [{\citenamefont {Disseler}\ \emph
  {et~al.}(2012{\natexlab{a}})\citenamefont {Disseler}, \citenamefont {Dhital},
  \citenamefont {Hogan}, \citenamefont {Amato}, \citenamefont {Giblin},
  \citenamefont {de~la Cruz}, \citenamefont {Daoud-Aladine}, \citenamefont
  {Wilson},\ and\ \citenamefont {Graf}}]{disseler2012magnetic1}%
  \BibitemOpen
  \bibfield  {author} {\bibinfo {author} {\bibfnamefont {S.~M.}\ \bibnamefont
  {Disseler}}, \bibinfo {author} {\bibfnamefont {C.}~\bibnamefont {Dhital}},
  \bibinfo {author} {\bibfnamefont {T.~C.}\ \bibnamefont {Hogan}}, \bibinfo
  {author} {\bibfnamefont {A.}~\bibnamefont {Amato}}, \bibinfo {author}
  {\bibfnamefont {S.~R.}\ \bibnamefont {Giblin}}, \bibinfo {author}
  {\bibfnamefont {C.}~\bibnamefont {de~la Cruz}}, \bibinfo {author}
  {\bibfnamefont {A.}~\bibnamefont {Daoud-Aladine}}, \bibinfo {author}
  {\bibfnamefont {S.~D.}\ \bibnamefont {Wilson}}, \ and\ \bibinfo {author}
  {\bibfnamefont {M.~J.}\ \bibnamefont {Graf}},\ }\href {\doibase
  10.1103/PhysRevB.85.174441} {\bibfield  {journal} {\bibinfo  {journal} {Phys.
  Rev. B}\ }\textbf {\bibinfo {volume} {85}},\ \bibinfo {pages} {174441}
  (\bibinfo {year} {2012}{\natexlab{a}})}\BibitemShut {NoStop}%
\bibitem [{\citenamefont {Disseler}\ \emph
  {et~al.}(2012{\natexlab{b}})\citenamefont {Disseler}, \citenamefont {Dhital},
  \citenamefont {Amato}, \citenamefont {Giblin}, \citenamefont {de~la Cruz},
  \citenamefont {Wilson},\ and\ \citenamefont {Graf}}]{disseler2012magnetic2}%
  \BibitemOpen
  \bibfield  {author} {\bibinfo {author} {\bibfnamefont {S.~M.}\ \bibnamefont
  {Disseler}}, \bibinfo {author} {\bibfnamefont {C.}~\bibnamefont {Dhital}},
  \bibinfo {author} {\bibfnamefont {A.}~\bibnamefont {Amato}}, \bibinfo
  {author} {\bibfnamefont {S.~R.}\ \bibnamefont {Giblin}}, \bibinfo {author}
  {\bibfnamefont {C.}~\bibnamefont {de~la Cruz}}, \bibinfo {author}
  {\bibfnamefont {S.~D.}\ \bibnamefont {Wilson}}, \ and\ \bibinfo {author}
  {\bibfnamefont {M.~J.}\ \bibnamefont {Graf}},\ }\href {\doibase
  10.1103/PhysRevB.86.014428} {\bibfield  {journal} {\bibinfo  {journal} {Phys.
  Rev. B}\ }\textbf {\bibinfo {volume} {86}},\ \bibinfo {pages} {014428}
  (\bibinfo {year} {2012}{\natexlab{b}})}\BibitemShut {NoStop}%
\bibitem [{\citenamefont {Graf}\ \emph {et~al.}(2014)\citenamefont {Graf},
  \citenamefont {Disseler}, \citenamefont {Dhital}, \citenamefont {Hogan},
  \citenamefont {Bojko}, \citenamefont {Amato}, \citenamefont {Luetkens},
  \citenamefont {Baines}, \citenamefont {Margineda}, \citenamefont {Giblin},
  \citenamefont {Jura},\ and\ \citenamefont {Wilson}}]{graf2014magnetism}%
  \BibitemOpen
  \bibfield  {author} {\bibinfo {author} {\bibfnamefont {M.~J.}\ \bibnamefont
  {Graf}}, \bibinfo {author} {\bibfnamefont {S.~M.}\ \bibnamefont {Disseler}},
  \bibinfo {author} {\bibfnamefont {C.}~\bibnamefont {Dhital}}, \bibinfo
  {author} {\bibfnamefont {T.}~\bibnamefont {Hogan}}, \bibinfo {author}
  {\bibfnamefont {M.}~\bibnamefont {Bojko}}, \bibinfo {author} {\bibfnamefont
  {A.}~\bibnamefont {Amato}}, \bibinfo {author} {\bibfnamefont
  {H.}~\bibnamefont {Luetkens}}, \bibinfo {author} {\bibfnamefont
  {C.}~\bibnamefont {Baines}}, \bibinfo {author} {\bibfnamefont
  {D.}~\bibnamefont {Margineda}}, \bibinfo {author} {\bibfnamefont {S.~R.}\
  \bibnamefont {Giblin}}, \bibinfo {author} {\bibfnamefont {M.}~\bibnamefont
  {Jura}}, \ and\ \bibinfo {author} {\bibfnamefont {S.~D.}\ \bibnamefont
  {Wilson}},\ }\href {http://stacks.iop.org/1742-6596/551/i=1/a=012020}
  {\bibfield  {journal} {\bibinfo  {journal} {J. Phys. Conf. Ser.}\ }\textbf
  {\bibinfo {volume} {551}},\ \bibinfo {pages} {012020} (\bibinfo {year}
  {2014})}\BibitemShut {NoStop}%
\bibitem [{\citenamefont {Tomiyasu}\ \emph {et~al.}(2012)\citenamefont
  {Tomiyasu}, \citenamefont {Matsuhira}, \citenamefont {Iwasa}, \citenamefont
  {Watahiki}, \citenamefont {Takagi}, \citenamefont {Wakeshima}, \citenamefont
  {Hinatsu}, \citenamefont {Yokoyama}, \citenamefont {Ohoyama},\ and\
  \citenamefont {Yamada}}]{tomiyasu2012emergence}%
  \BibitemOpen
  \bibfield  {author} {\bibinfo {author} {\bibfnamefont {K.}~\bibnamefont
  {Tomiyasu}}, \bibinfo {author} {\bibfnamefont {K.}~\bibnamefont {Matsuhira}},
  \bibinfo {author} {\bibfnamefont {K.}~\bibnamefont {Iwasa}}, \bibinfo
  {author} {\bibfnamefont {M.}~\bibnamefont {Watahiki}}, \bibinfo {author}
  {\bibfnamefont {S.}~\bibnamefont {Takagi}}, \bibinfo {author} {\bibfnamefont
  {M.}~\bibnamefont {Wakeshima}}, \bibinfo {author} {\bibfnamefont
  {Y.}~\bibnamefont {Hinatsu}}, \bibinfo {author} {\bibfnamefont
  {M.}~\bibnamefont {Yokoyama}}, \bibinfo {author} {\bibfnamefont
  {K.}~\bibnamefont {Ohoyama}}, \ and\ \bibinfo {author} {\bibfnamefont
  {K.}~\bibnamefont {Yamada}},\ }\href {\doibase 10.1143/JPSJ.81.034709}
  {\bibfield  {journal} {\bibinfo  {journal} {J. Phys. Soc. Jpn.}\ }\textbf
  {\bibinfo {volume} {81}},\ \bibinfo {pages} {034709} (\bibinfo {year}
  {2012})}\BibitemShut {NoStop}%
\bibitem [{\citenamefont {Lefran\ifmmode~\mbox{\c{c}}\else \c{c}\fi{}ois}\
  \emph {et~al.}(2015)\citenamefont {Lefran\ifmmode~\mbox{\c{c}}\else
  \c{c}\fi{}ois}, \citenamefont {Simonet}, \citenamefont {Ballou},
  \citenamefont {Lhotel}, \citenamefont {Hadj-Azzem}, \citenamefont
  {Kodjikian}, \citenamefont {Lejay}, \citenamefont {Manuel}, \citenamefont
  {Khalyavin},\ and\ \citenamefont {Chapon}}]{lefrancois2015anisotropy}%
  \BibitemOpen
  \bibfield  {author} {\bibinfo {author} {\bibfnamefont {E.}~\bibnamefont
  {Lefran\ifmmode~\mbox{\c{c}}\else \c{c}\fi{}ois}}, \bibinfo {author}
  {\bibfnamefont {V.}~\bibnamefont {Simonet}}, \bibinfo {author} {\bibfnamefont
  {R.}~\bibnamefont {Ballou}}, \bibinfo {author} {\bibfnamefont
  {E.}~\bibnamefont {Lhotel}}, \bibinfo {author} {\bibfnamefont
  {A.}~\bibnamefont {Hadj-Azzem}}, \bibinfo {author} {\bibfnamefont
  {S.}~\bibnamefont {Kodjikian}}, \bibinfo {author} {\bibfnamefont
  {P.}~\bibnamefont {Lejay}}, \bibinfo {author} {\bibfnamefont
  {P.}~\bibnamefont {Manuel}}, \bibinfo {author} {\bibfnamefont
  {D.}~\bibnamefont {Khalyavin}}, \ and\ \bibinfo {author} {\bibfnamefont
  {L.~C.}\ \bibnamefont {Chapon}},\ }\href {\doibase
  10.1103/PhysRevLett.114.247202} {\bibfield  {journal} {\bibinfo  {journal}
  {Phys. Rev. Lett.}\ }\textbf {\bibinfo {volume} {114}},\ \bibinfo {pages}
  {247202} (\bibinfo {year} {2015})}\BibitemShut {NoStop}%
\bibitem [{\citenamefont {Guo}\ \emph {et~al.}(2016)\citenamefont {Guo},
  \citenamefont {Ritter},\ and\ \citenamefont {Komarek}}]{guo2016direct}%
  \BibitemOpen
  \bibfield  {author} {\bibinfo {author} {\bibfnamefont {H.}~\bibnamefont
  {Guo}}, \bibinfo {author} {\bibfnamefont {C.}~\bibnamefont {Ritter}}, \ and\
  \bibinfo {author} {\bibfnamefont {A.~C.}\ \bibnamefont {Komarek}},\ }\href
  {\doibase 10.1103/PhysRevB.94.161102} {\bibfield  {journal} {\bibinfo
  {journal} {Phys. Rev. B}\ }\textbf {\bibinfo {volume} {94}},\ \bibinfo
  {pages} {161102} (\bibinfo {year} {2016})}\BibitemShut {NoStop}%
\bibitem [{\citenamefont {Guo}\ \emph {et~al.}(2017)\citenamefont {Guo},
  \citenamefont {Ritter},\ and\ \citenamefont {Komarek}}]{guo2017magnetic}%
  \BibitemOpen
  \bibfield  {author} {\bibinfo {author} {\bibfnamefont {H.}~\bibnamefont
  {Guo}}, \bibinfo {author} {\bibfnamefont {C.}~\bibnamefont {Ritter}}, \ and\
  \bibinfo {author} {\bibfnamefont {A.~C.}\ \bibnamefont {Komarek}},\ }\href
  {\doibase 10.1103/PhysRevB.96.144415} {\bibfield  {journal} {\bibinfo
  {journal} {Phys. Rev. B}\ }\textbf {\bibinfo {volume} {96}},\ \bibinfo
  {pages} {144415} (\bibinfo {year} {2017})}\BibitemShut {NoStop}%
\bibitem [{\citenamefont {Sagayama}\ \emph {et~al.}(2013)\citenamefont
  {Sagayama}, \citenamefont {Uematsu}, \citenamefont {Arima}, \citenamefont
  {Sugimoto}, \citenamefont {Ishikawa}, \citenamefont {O'Farrell},\ and\
  \citenamefont {Nakatsuji}}]{sagayama2013determination}%
  \BibitemOpen
  \bibfield  {author} {\bibinfo {author} {\bibfnamefont {H.}~\bibnamefont
  {Sagayama}}, \bibinfo {author} {\bibfnamefont {D.}~\bibnamefont {Uematsu}},
  \bibinfo {author} {\bibfnamefont {T.}~\bibnamefont {Arima}}, \bibinfo
  {author} {\bibfnamefont {K.}~\bibnamefont {Sugimoto}}, \bibinfo {author}
  {\bibfnamefont {J.~J.}\ \bibnamefont {Ishikawa}}, \bibinfo {author}
  {\bibfnamefont {E.}~\bibnamefont {O'Farrell}}, \ and\ \bibinfo {author}
  {\bibfnamefont {S.}~\bibnamefont {Nakatsuji}},\ }\href {\doibase
  10.1103/PhysRevB.87.100403} {\bibfield  {journal} {\bibinfo  {journal} {Phys.
  Rev. B}\ }\textbf {\bibinfo {volume} {87}},\ \bibinfo {pages} {100403}
  (\bibinfo {year} {2013})}\BibitemShut {NoStop}%
\bibitem [{\citenamefont {Clancy}\ \emph {et~al.}(2016)\citenamefont {Clancy},
  \citenamefont {Gretarsson}, \citenamefont {Lee}, \citenamefont {Tian},
  \citenamefont {Kim}, \citenamefont {Upton}, \citenamefont {Casa},
  \citenamefont {Gog}, \citenamefont {Islam}, \citenamefont {Jeon},
  \citenamefont {Kim}, \citenamefont {Desgreniers}, \citenamefont {Kim},
  \citenamefont {Julian},\ and\ \citenamefont {Kim}}]{clancy2016x-ray}%
  \BibitemOpen
  \bibfield  {author} {\bibinfo {author} {\bibfnamefont {J.~P.}\ \bibnamefont
  {Clancy}}, \bibinfo {author} {\bibfnamefont {H.}~\bibnamefont {Gretarsson}},
  \bibinfo {author} {\bibfnamefont {E.~K.~H.}\ \bibnamefont {Lee}}, \bibinfo
  {author} {\bibfnamefont {D.}~\bibnamefont {Tian}}, \bibinfo {author}
  {\bibfnamefont {J.}~\bibnamefont {Kim}}, \bibinfo {author} {\bibfnamefont
  {M.~H.}\ \bibnamefont {Upton}}, \bibinfo {author} {\bibfnamefont
  {D.}~\bibnamefont {Casa}}, \bibinfo {author} {\bibfnamefont {T.}~\bibnamefont
  {Gog}}, \bibinfo {author} {\bibfnamefont {Z.}~\bibnamefont {Islam}}, \bibinfo
  {author} {\bibfnamefont {B.-G.}\ \bibnamefont {Jeon}}, \bibinfo {author}
  {\bibfnamefont {K.~H.}\ \bibnamefont {Kim}}, \bibinfo {author} {\bibfnamefont
  {S.}~\bibnamefont {Desgreniers}}, \bibinfo {author} {\bibfnamefont {Y.~B.}\
  \bibnamefont {Kim}}, \bibinfo {author} {\bibfnamefont {S.~J.}\ \bibnamefont
  {Julian}}, \ and\ \bibinfo {author} {\bibfnamefont {Y.-J.}\ \bibnamefont
  {Kim}},\ }\href {\doibase 10.1103/PhysRevB.94.024408} {\bibfield  {journal}
  {\bibinfo  {journal} {Phys. Rev. B}\ }\textbf {\bibinfo {volume} {94}},\
  \bibinfo {pages} {024408} (\bibinfo {year} {2016})}\BibitemShut {NoStop}%
\bibitem [{\citenamefont {Donnerer}\ \emph {et~al.}(2016)\citenamefont
  {Donnerer}, \citenamefont {Rahn}, \citenamefont {Sala}, \citenamefont {Vale},
  \citenamefont {Pincini}, \citenamefont {Strempfer}, \citenamefont {Krisch},
  \citenamefont {Prabhakaran}, \citenamefont {Boothroyd},\ and\ \citenamefont
  {McMorrow}}]{donnerer2016all-in}%
  \BibitemOpen
  \bibfield  {author} {\bibinfo {author} {\bibfnamefont {C.}~\bibnamefont
  {Donnerer}}, \bibinfo {author} {\bibfnamefont {M.~C.}\ \bibnamefont {Rahn}},
  \bibinfo {author} {\bibfnamefont {M.~M.}\ \bibnamefont {Sala}}, \bibinfo
  {author} {\bibfnamefont {J.~G.}\ \bibnamefont {Vale}}, \bibinfo {author}
  {\bibfnamefont {D.}~\bibnamefont {Pincini}}, \bibinfo {author} {\bibfnamefont
  {J.}~\bibnamefont {Strempfer}}, \bibinfo {author} {\bibfnamefont
  {M.}~\bibnamefont {Krisch}}, \bibinfo {author} {\bibfnamefont
  {D.}~\bibnamefont {Prabhakaran}}, \bibinfo {author} {\bibfnamefont {A.~T.}\
  \bibnamefont {Boothroyd}}, \ and\ \bibinfo {author} {\bibfnamefont {D.~F.}\
  \bibnamefont {McMorrow}},\ }\href {\doibase 10.1103/PhysRevLett.117.037201}
  {\bibfield  {journal} {\bibinfo  {journal} {Phys. Rev. Lett.}\ }\textbf
  {\bibinfo {volume} {117}},\ \bibinfo {pages} {037201} (\bibinfo {year}
  {2016})}\BibitemShut {NoStop}%
\bibitem [{\citenamefont {Ueda}\ \emph {et~al.}(2016)\citenamefont {Ueda},
  \citenamefont {Fujioka},\ and\ \citenamefont {Tokura}}]{ueda2016variation}%
  \BibitemOpen
  \bibfield  {author} {\bibinfo {author} {\bibfnamefont {K.}~\bibnamefont
  {Ueda}}, \bibinfo {author} {\bibfnamefont {J.}~\bibnamefont {Fujioka}}, \
  and\ \bibinfo {author} {\bibfnamefont {Y.}~\bibnamefont {Tokura}},\ }\href
  {\doibase 10.1103/PhysRevB.93.245120} {\bibfield  {journal} {\bibinfo
  {journal} {Phys. Rev. B}\ }\textbf {\bibinfo {volume} {93}},\ \bibinfo
  {pages} {245120} (\bibinfo {year} {2016})}\BibitemShut {NoStop}%
\bibitem [{\citenamefont {Gardner}\ \emph {et~al.}(2010)\citenamefont
  {Gardner}, \citenamefont {Gingras},\ and\ \citenamefont
  {Greedan}}]{gardner2010magnetic}%
  \BibitemOpen
  \bibfield  {author} {\bibinfo {author} {\bibfnamefont {J.~S.}\ \bibnamefont
  {Gardner}}, \bibinfo {author} {\bibfnamefont {M.~J.~P.}\ \bibnamefont
  {Gingras}}, \ and\ \bibinfo {author} {\bibfnamefont {J.~E.}\ \bibnamefont
  {Greedan}},\ }\href {\doibase 10.1103/RevModPhys.82.53} {\bibfield  {journal}
  {\bibinfo  {journal} {Rev. Mod. Phys.}\ }\textbf {\bibinfo {volume} {82}},\
  \bibinfo {pages} {53} (\bibinfo {year} {2010})}\BibitemShut {NoStop}%
\bibitem [{\citenamefont {MacLaughlin}\ \emph {et~al.}(2015)\citenamefont
  {MacLaughlin}, \citenamefont {Bernal}, \citenamefont {Shu}, \citenamefont
  {Ishikawa}, \citenamefont {Matsumoto}, \citenamefont {Wen}, \citenamefont
  {Mourigal}, \citenamefont {Stock}, \citenamefont {Ehlers}, \citenamefont
  {Broholm}, \citenamefont {Machida}, \citenamefont {Kimura}, \citenamefont
  {Nakatsuji}, \citenamefont {Shimura},\ and\ \citenamefont
  {Sakakibara}}]{MacLaughlin2015unstable}%
  \BibitemOpen
  \bibfield  {author} {\bibinfo {author} {\bibfnamefont {D.~E.}\ \bibnamefont
  {MacLaughlin}}, \bibinfo {author} {\bibfnamefont {O.~O.}\ \bibnamefont
  {Bernal}}, \bibinfo {author} {\bibfnamefont {L.}~\bibnamefont {Shu}},
  \bibinfo {author} {\bibfnamefont {J.}~\bibnamefont {Ishikawa}}, \bibinfo
  {author} {\bibfnamefont {Y.}~\bibnamefont {Matsumoto}}, \bibinfo {author}
  {\bibfnamefont {J.-J.}\ \bibnamefont {Wen}}, \bibinfo {author} {\bibfnamefont
  {M.}~\bibnamefont {Mourigal}}, \bibinfo {author} {\bibfnamefont
  {C.}~\bibnamefont {Stock}}, \bibinfo {author} {\bibfnamefont
  {G.}~\bibnamefont {Ehlers}}, \bibinfo {author} {\bibfnamefont {C.~L.}\
  \bibnamefont {Broholm}}, \bibinfo {author} {\bibfnamefont {Y.}~\bibnamefont
  {Machida}}, \bibinfo {author} {\bibfnamefont {K.}~\bibnamefont {Kimura}},
  \bibinfo {author} {\bibfnamefont {S.}~\bibnamefont {Nakatsuji}}, \bibinfo
  {author} {\bibfnamefont {Y.}~\bibnamefont {Shimura}}, \ and\ \bibinfo
  {author} {\bibfnamefont {T.}~\bibnamefont {Sakakibara}},\ }\href {\doibase
  10.1103/PhysRevB.92.054432} {\bibfield  {journal} {\bibinfo  {journal} {Phys.
  Rev. B}\ }\textbf {\bibinfo {volume} {92}},\ \bibinfo {pages} {054432}
  (\bibinfo {year} {2015})}\BibitemShut {NoStop}%
\bibitem [{\citenamefont {Telang}\ \emph {et~al.}(2018)\citenamefont {Telang},
  \citenamefont {Mishra}, \citenamefont {Sood},\ and\ \citenamefont
  {Singh}}]{Telang2018dilute}%
  \BibitemOpen
  \bibfield  {author} {\bibinfo {author} {\bibfnamefont {P.}~\bibnamefont
  {Telang}}, \bibinfo {author} {\bibfnamefont {K.}~\bibnamefont {Mishra}},
  \bibinfo {author} {\bibfnamefont {A.~K.}\ \bibnamefont {Sood}}, \ and\
  \bibinfo {author} {\bibfnamefont {S.}~\bibnamefont {Singh}},\ }\href
  {\doibase 10.1103/PhysRevB.97.235118} {\bibfield  {journal} {\bibinfo
  {journal} {Phys. Rev. B}\ }\textbf {\bibinfo {volume} {97}},\ \bibinfo
  {pages} {235118} (\bibinfo {year} {2018})}\BibitemShut {NoStop}%
\bibitem [{\citenamefont {Nakayama}\ \emph {et~al.}(2016)\citenamefont
  {Nakayama}, \citenamefont {Kondo}, \citenamefont {Tian}, \citenamefont
  {Ishikawa}, \citenamefont {Halim}, \citenamefont {Bareille}, \citenamefont
  {Malaeb}, \citenamefont {Kuroda}, \citenamefont {Tomita}, \citenamefont
  {Ideta}, \citenamefont {Tanaka}, \citenamefont {Matsunami}, \citenamefont
  {Kimura}, \citenamefont {Inami}, \citenamefont {Ono}, \citenamefont
  {Kumigashira}, \citenamefont {Balents}, \citenamefont {Nakatsuji},\ and\
  \citenamefont {Shin}}]{nakayama2016slater}%
  \BibitemOpen
  \bibfield  {author} {\bibinfo {author} {\bibfnamefont {M.}~\bibnamefont
  {Nakayama}}, \bibinfo {author} {\bibfnamefont {T.}~\bibnamefont {Kondo}},
  \bibinfo {author} {\bibfnamefont {Z.}~\bibnamefont {Tian}}, \bibinfo {author}
  {\bibfnamefont {J.~J.}\ \bibnamefont {Ishikawa}}, \bibinfo {author}
  {\bibfnamefont {M.}~\bibnamefont {Halim}}, \bibinfo {author} {\bibfnamefont
  {C.}~\bibnamefont {Bareille}}, \bibinfo {author} {\bibfnamefont
  {W.}~\bibnamefont {Malaeb}}, \bibinfo {author} {\bibfnamefont
  {K.}~\bibnamefont {Kuroda}}, \bibinfo {author} {\bibfnamefont
  {T.}~\bibnamefont {Tomita}}, \bibinfo {author} {\bibfnamefont
  {S.}~\bibnamefont {Ideta}}, \bibinfo {author} {\bibfnamefont
  {K.}~\bibnamefont {Tanaka}}, \bibinfo {author} {\bibfnamefont
  {M.}~\bibnamefont {Matsunami}}, \bibinfo {author} {\bibfnamefont
  {S.}~\bibnamefont {Kimura}}, \bibinfo {author} {\bibfnamefont
  {N.}~\bibnamefont {Inami}}, \bibinfo {author} {\bibfnamefont
  {K.}~\bibnamefont {Ono}}, \bibinfo {author} {\bibfnamefont {H.}~\bibnamefont
  {Kumigashira}}, \bibinfo {author} {\bibfnamefont {L.}~\bibnamefont
  {Balents}}, \bibinfo {author} {\bibfnamefont {S.}~\bibnamefont {Nakatsuji}},
  \ and\ \bibinfo {author} {\bibfnamefont {S.}~\bibnamefont {Shin}},\ }\href
  {\doibase 10.1103/PhysRevLett.117.056403} {\bibfield  {journal} {\bibinfo
  {journal} {Phys. Rev. Lett.}\ }\textbf {\bibinfo {volume} {117}},\ \bibinfo
  {pages} {056403} (\bibinfo {year} {2016})}\BibitemShut {NoStop}%
\bibitem [{\citenamefont {Matsuhira}\ \emph {et~al.}(2011)\citenamefont
  {Matsuhira}, \citenamefont {Wakeshima}, \citenamefont {Hinatsu},\ and\
  \citenamefont {Takagi}}]{matsuhira2011metal}%
  \BibitemOpen
  \bibfield  {author} {\bibinfo {author} {\bibfnamefont {K.}~\bibnamefont
  {Matsuhira}}, \bibinfo {author} {\bibfnamefont {M.}~\bibnamefont
  {Wakeshima}}, \bibinfo {author} {\bibfnamefont {Y.}~\bibnamefont {Hinatsu}},
  \ and\ \bibinfo {author} {\bibfnamefont {S.}~\bibnamefont {Takagi}},\ }\href
  {\doibase 10.1143/JPSJ.80.094701} {\bibfield  {journal} {\bibinfo  {journal}
  {J. Phys. Soc. Jpn.}\ }\textbf {\bibinfo {volume} {80}},\ \bibinfo {pages}
  {094701} (\bibinfo {year} {2011})}\BibitemShut {NoStop}%
\bibitem [{\citenamefont {Millican}\ \emph {et~al.}(2007)\citenamefont
  {Millican}, \citenamefont {Macaluso}, \citenamefont {Nakatsuji},
  \citenamefont {Machida}, \citenamefont {Maeno},\ and\ \citenamefont
  {Chan}}]{Millican2007crystal}%
  \BibitemOpen
  \bibfield  {author} {\bibinfo {author} {\bibfnamefont {J.~N.}\ \bibnamefont
  {Millican}}, \bibinfo {author} {\bibfnamefont {R.~T.}\ \bibnamefont
  {Macaluso}}, \bibinfo {author} {\bibfnamefont {S.}~\bibnamefont {Nakatsuji}},
  \bibinfo {author} {\bibfnamefont {Y.}~\bibnamefont {Machida}}, \bibinfo
  {author} {\bibfnamefont {Y.}~\bibnamefont {Maeno}}, \ and\ \bibinfo {author}
  {\bibfnamefont {J.~Y.}\ \bibnamefont {Chan}},\ }\href {\doibase
  http://dx.doi.org/10.1016/j.materresbull.2006.08.011} {\bibfield  {journal}
  {\bibinfo  {journal} {Mater. Res. Bull.}\ }\textbf {\bibinfo {volume} {42}},\
  \bibinfo {pages} {928 } (\bibinfo {year} {2007})}\BibitemShut {NoStop}%
\bibitem [{\citenamefont {Dmitrienko}\ \emph {et~al.}(2005)\citenamefont
  {Dmitrienko}, \citenamefont {Ishida}, \citenamefont {Kirfel},\ and\
  \citenamefont {Ovchinnikova}}]{dmitrienko2005polarization}%
  \BibitemOpen
  \bibfield  {author} {\bibinfo {author} {\bibfnamefont {V.~E.}\ \bibnamefont
  {Dmitrienko}}, \bibinfo {author} {\bibfnamefont {K.}~\bibnamefont {Ishida}},
  \bibinfo {author} {\bibfnamefont {A.}~\bibnamefont {Kirfel}}, \ and\ \bibinfo
  {author} {\bibfnamefont {E.~N.}\ \bibnamefont {Ovchinnikova}},\ }\href
  {\doibase 10.1107/S0108767305018209} {\bibfield  {journal} {\bibinfo
  {journal} {Acta Crystallogr. Sect. A}\ }\textbf {\bibinfo {volume} {61}},\
  \bibinfo {pages} {481} (\bibinfo {year} {2005})}\BibitemShut {NoStop}%
\bibitem [{\citenamefont {Donnerer}(2018)}]{donnerer2018magnetic}%
  \BibitemOpen
  \bibfield  {author} {\bibinfo {author} {\bibfnamefont {C.}~\bibnamefont
  {Donnerer}},\ }\emph {\bibinfo {title} {X-ray studies of magnetic and
  structural transitions in iridates}},\ \href
  {http://discovery.ucl.ac.uk/10047266/} {Ph.D. thesis},\ \bibinfo  {school}
  {University College London} (\bibinfo {year} {2018})\BibitemShut {NoStop}%
\bibitem [{\citenamefont {Abragam}\ and\ \citenamefont
  {Bleaney}(1970)}]{abragam1970electron}%
  \BibitemOpen
  \bibfield  {author} {\bibinfo {author} {\bibfnamefont {A.}~\bibnamefont
  {Abragam}}\ and\ \bibinfo {author} {\bibfnamefont {B.}~\bibnamefont
  {Bleaney}},\ }\href@noop {} {\emph {\bibinfo {title} {Electron paramagnetic
  resonance of transition ions}}},\ International series of monographs on
  physics\ (\bibinfo  {publisher} {Clarendon P.},\ \bibinfo {year}
  {1970})\BibitemShut {NoStop}%
\bibitem [{\citenamefont {Jackeli}\ and\ \citenamefont
  {Khaliullin}(2009)}]{jackeli2009mott}%
  \BibitemOpen
  \bibfield  {author} {\bibinfo {author} {\bibfnamefont {G.}~\bibnamefont
  {Jackeli}}\ and\ \bibinfo {author} {\bibfnamefont {G.}~\bibnamefont
  {Khaliullin}},\ }\href {\doibase 10.1103/PhysRevLett.102.017205} {\bibfield
  {journal} {\bibinfo  {journal} {Phys. Rev. Lett.}\ }\textbf {\bibinfo
  {volume} {102}},\ \bibinfo {pages} {017205} (\bibinfo {year}
  {2009})}\BibitemShut {NoStop}%
\bibitem [{\citenamefont {Ament}\ \emph {et~al.}(2011)\citenamefont {Ament},
  \citenamefont {Khaliullin},\ and\ \citenamefont {van~den
  Brink}}]{ament2011theory}%
  \BibitemOpen
  \bibfield  {author} {\bibinfo {author} {\bibfnamefont {L.~J.~P.}\
  \bibnamefont {Ament}}, \bibinfo {author} {\bibfnamefont {G.}~\bibnamefont
  {Khaliullin}}, \ and\ \bibinfo {author} {\bibfnamefont {J.}~\bibnamefont
  {van~den Brink}},\ }\href {\doibase 10.1103/PhysRevB.84.020403} {\bibfield
  {journal} {\bibinfo  {journal} {Phys. Rev. B}\ }\textbf {\bibinfo {volume}
  {84}},\ \bibinfo {pages} {020403} (\bibinfo {year} {2011})}\BibitemShut
  {NoStop}%
\bibitem [{\citenamefont {Liu}\ \emph {et~al.}(2012)\citenamefont {Liu},
  \citenamefont {Katukuri}, \citenamefont {Hozoi}, \citenamefont {Yin},
  \citenamefont {Dean}, \citenamefont {Upton}, \citenamefont {Kim},
  \citenamefont {Casa}, \citenamefont {Said}, \citenamefont {Gog},
  \citenamefont {Qi}, \citenamefont {Cao}, \citenamefont {Tsvelik},
  \citenamefont {van~den Brink},\ and\ \citenamefont {Hill}}]{liu2012testing}%
  \BibitemOpen
  \bibfield  {author} {\bibinfo {author} {\bibfnamefont {X.}~\bibnamefont
  {Liu}}, \bibinfo {author} {\bibfnamefont {V.~M.}\ \bibnamefont {Katukuri}},
  \bibinfo {author} {\bibfnamefont {L.}~\bibnamefont {Hozoi}}, \bibinfo
  {author} {\bibfnamefont {W.-G.}\ \bibnamefont {Yin}}, \bibinfo {author}
  {\bibfnamefont {M.~P.~M.}\ \bibnamefont {Dean}}, \bibinfo {author}
  {\bibfnamefont {M.~H.}\ \bibnamefont {Upton}}, \bibinfo {author}
  {\bibfnamefont {J.}~\bibnamefont {Kim}}, \bibinfo {author} {\bibfnamefont
  {D.}~\bibnamefont {Casa}}, \bibinfo {author} {\bibfnamefont {A.}~\bibnamefont
  {Said}}, \bibinfo {author} {\bibfnamefont {T.}~\bibnamefont {Gog}}, \bibinfo
  {author} {\bibfnamefont {T.~F.}\ \bibnamefont {Qi}}, \bibinfo {author}
  {\bibfnamefont {G.}~\bibnamefont {Cao}}, \bibinfo {author} {\bibfnamefont
  {A.~M.}\ \bibnamefont {Tsvelik}}, \bibinfo {author} {\bibfnamefont
  {J.}~\bibnamefont {van~den Brink}}, \ and\ \bibinfo {author} {\bibfnamefont
  {J.~P.}\ \bibnamefont {Hill}},\ }\href {\doibase
  10.1103/PhysRevLett.109.157401} {\bibfield  {journal} {\bibinfo  {journal}
  {Phys. Rev. Lett.}\ }\textbf {\bibinfo {volume} {109}},\ \bibinfo {pages}
  {157401} (\bibinfo {year} {2012})}\BibitemShut {NoStop}%
\bibitem [{\citenamefont {Ohgushi}\ \emph {et~al.}(2013)\citenamefont
  {Ohgushi}, \citenamefont {Yamaura}, \citenamefont {Ohsumi}, \citenamefont
  {Sugimoto}, \citenamefont {Takeshita}, \citenamefont {Tokuda}, \citenamefont
  {Takagi}, \citenamefont {Takata},\ and\ \citenamefont
  {Arima}}]{ohgushi2013resonant}%
  \BibitemOpen
  \bibfield  {author} {\bibinfo {author} {\bibfnamefont {K.}~\bibnamefont
  {Ohgushi}}, \bibinfo {author} {\bibfnamefont {J.-i.}\ \bibnamefont
  {Yamaura}}, \bibinfo {author} {\bibfnamefont {H.}~\bibnamefont {Ohsumi}},
  \bibinfo {author} {\bibfnamefont {K.}~\bibnamefont {Sugimoto}}, \bibinfo
  {author} {\bibfnamefont {S.}~\bibnamefont {Takeshita}}, \bibinfo {author}
  {\bibfnamefont {A.}~\bibnamefont {Tokuda}}, \bibinfo {author} {\bibfnamefont
  {H.}~\bibnamefont {Takagi}}, \bibinfo {author} {\bibfnamefont
  {M.}~\bibnamefont {Takata}}, \ and\ \bibinfo {author} {\bibfnamefont {T.-h.}\
  \bibnamefont {Arima}},\ }\href {\doibase 10.1103/PhysRevLett.110.217212}
  {\bibfield  {journal} {\bibinfo  {journal} {Phys. Rev. Lett.}\ }\textbf
  {\bibinfo {volume} {110}},\ \bibinfo {pages} {217212} (\bibinfo {year}
  {2013})}\BibitemShut {NoStop}%
\bibitem [{\citenamefont {Hozoi}\ \emph {et~al.}(2014)\citenamefont {Hozoi},
  \citenamefont {Gretarsson}, \citenamefont {Clancy}, \citenamefont {Jeon},
  \citenamefont {Lee}, \citenamefont {Kim}, \citenamefont {Yushankhai},
  \citenamefont {Fulde}, \citenamefont {Casa}, \citenamefont {Gog},
  \citenamefont {Kim}, \citenamefont {Said}, \citenamefont {Upton},
  \citenamefont {Kim},\ and\ \citenamefont {van~den Brink}}]{hozoi2014longer}%
  \BibitemOpen
  \bibfield  {author} {\bibinfo {author} {\bibfnamefont {L.}~\bibnamefont
  {Hozoi}}, \bibinfo {author} {\bibfnamefont {H.}~\bibnamefont {Gretarsson}},
  \bibinfo {author} {\bibfnamefont {J.~P.}\ \bibnamefont {Clancy}}, \bibinfo
  {author} {\bibfnamefont {B.-G.}\ \bibnamefont {Jeon}}, \bibinfo {author}
  {\bibfnamefont {B.}~\bibnamefont {Lee}}, \bibinfo {author} {\bibfnamefont
  {K.~H.}\ \bibnamefont {Kim}}, \bibinfo {author} {\bibfnamefont
  {V.}~\bibnamefont {Yushankhai}}, \bibinfo {author} {\bibfnamefont
  {P.}~\bibnamefont {Fulde}}, \bibinfo {author} {\bibfnamefont
  {D.}~\bibnamefont {Casa}}, \bibinfo {author} {\bibfnamefont {T.}~\bibnamefont
  {Gog}}, \bibinfo {author} {\bibfnamefont {J.}~\bibnamefont {Kim}}, \bibinfo
  {author} {\bibfnamefont {A.~H.}\ \bibnamefont {Said}}, \bibinfo {author}
  {\bibfnamefont {M.~H.}\ \bibnamefont {Upton}}, \bibinfo {author}
  {\bibfnamefont {Y.-J.}\ \bibnamefont {Kim}}, \ and\ \bibinfo {author}
  {\bibfnamefont {J.}~\bibnamefont {van~den Brink}},\ }\href {\doibase
  10.1103/PhysRevB.89.115111} {\bibfield  {journal} {\bibinfo  {journal} {Phys.
  Rev. B}\ }\textbf {\bibinfo {volume} {89}},\ \bibinfo {pages} {115111}
  (\bibinfo {year} {2014})}\BibitemShut {NoStop}%
\bibitem [{\citenamefont {Moretti~Sala}\ \emph {et~al.}(2014)\citenamefont
  {Moretti~Sala}, \citenamefont {Ohgushi}, \citenamefont {Al-Zein},
  \citenamefont {Hirata}, \citenamefont {Monaco},\ and\ \citenamefont
  {Krisch}}]{moretti2014cairo}%
  \BibitemOpen
  \bibfield  {author} {\bibinfo {author} {\bibfnamefont {M.}~\bibnamefont
  {Moretti~Sala}}, \bibinfo {author} {\bibfnamefont {K.}~\bibnamefont
  {Ohgushi}}, \bibinfo {author} {\bibfnamefont {A.}~\bibnamefont {Al-Zein}},
  \bibinfo {author} {\bibfnamefont {Y.}~\bibnamefont {Hirata}}, \bibinfo
  {author} {\bibfnamefont {G.}~\bibnamefont {Monaco}}, \ and\ \bibinfo {author}
  {\bibfnamefont {M.}~\bibnamefont {Krisch}},\ }\href {\doibase
  10.1103/PhysRevLett.112.176402} {\bibfield  {journal} {\bibinfo  {journal}
  {Phys. Rev. Lett.}\ }\textbf {\bibinfo {volume} {112}},\ \bibinfo {pages}
  {176402} (\bibinfo {year} {2014})}\BibitemShut {NoStop}%
\bibitem [{Note1()}]{Note1}%
  \BibitemOpen
  \bibinfo {note} {The reason why we do not correct $I_{ats}$ for leakage is
  that (a) due to the stronger ATS signal this will comprise a much smaller
  relative error and (b) at the azimuthal angle where $I_{ats}^{\sigma \pi
  ^{\prime }}$ is maximised, the corresponding signal in $I_{ats}^{\sigma
  \sigma ^{\prime }}$ is minimised, which should substantially reduce the
  leakage.}\BibitemShut {Stop}%
\bibitem [{\citenamefont {Wang}\ \emph {et~al.}(2017)\citenamefont {Wang},
  \citenamefont {Go},\ and\ \citenamefont {Millis}}]{wang2017weyl}%
  \BibitemOpen
  \bibfield  {author} {\bibinfo {author} {\bibfnamefont {R.}~\bibnamefont
  {Wang}}, \bibinfo {author} {\bibfnamefont {A.}~\bibnamefont {Go}}, \ and\
  \bibinfo {author} {\bibfnamefont {A.}~\bibnamefont {Millis}},\ }\href
  {\doibase 10.1103/PhysRevB.96.195158} {\bibfield  {journal} {\bibinfo
  {journal} {Phys. Rev. B}\ }\textbf {\bibinfo {volume} {96}},\ \bibinfo
  {pages} {195158} (\bibinfo {year} {2017})}\BibitemShut {NoStop}%
\bibitem [{\citenamefont {Goedkoop}\ \emph {et~al.}(1988)\citenamefont
  {Goedkoop}, \citenamefont {Thole}, \citenamefont {van~der Laan},
  \citenamefont {Sawatzky}, \citenamefont {de~Groot},\ and\ \citenamefont
  {Fuggle}}]{goedkoop1988calculations}%
  \BibitemOpen
  \bibfield  {author} {\bibinfo {author} {\bibfnamefont {J.~B.}\ \bibnamefont
  {Goedkoop}}, \bibinfo {author} {\bibfnamefont {B.~T.}\ \bibnamefont {Thole}},
  \bibinfo {author} {\bibfnamefont {G.}~\bibnamefont {van~der Laan}}, \bibinfo
  {author} {\bibfnamefont {G.~A.}\ \bibnamefont {Sawatzky}}, \bibinfo {author}
  {\bibfnamefont {F.~M.~F.}\ \bibnamefont {de~Groot}}, \ and\ \bibinfo {author}
  {\bibfnamefont {J.~C.}\ \bibnamefont {Fuggle}},\ }\href {\doibase
  10.1103/PhysRevB.37.2086} {\bibfield  {journal} {\bibinfo  {journal} {Phys.
  Rev. B}\ }\textbf {\bibinfo {volume} {37}},\ \bibinfo {pages} {2086}
  (\bibinfo {year} {1988})}\BibitemShut {NoStop}%
\bibitem [{\citenamefont {Hill}\ and\ \citenamefont
  {McMorrow}(1996)}]{hill1996resonant}%
  \BibitemOpen
  \bibfield  {author} {\bibinfo {author} {\bibfnamefont {J.~P.}\ \bibnamefont
  {Hill}}\ and\ \bibinfo {author} {\bibfnamefont {D.~F.}\ \bibnamefont
  {McMorrow}},\ }\href {\doibase 10.1107/S0108767395012670} {\bibfield
  {journal} {\bibinfo  {journal} {Acta Crystallogr. Sect. A}\ }\textbf
  {\bibinfo {volume} {52}},\ \bibinfo {pages} {236} (\bibinfo {year}
  {1996})}\BibitemShut {NoStop}%
\bibitem [{\citenamefont {Collins}\ \emph {et~al.}(2001)\citenamefont
  {Collins}, \citenamefont {Laundy},\ and\ \citenamefont
  {Stunault}}]{collins2001anisotropic}%
  \BibitemOpen
  \bibfield  {author} {\bibinfo {author} {\bibfnamefont {S.~P.}\ \bibnamefont
  {Collins}}, \bibinfo {author} {\bibfnamefont {D.}~\bibnamefont {Laundy}}, \
  and\ \bibinfo {author} {\bibfnamefont {A.}~\bibnamefont {Stunault}},\ }\href
  {http://stacks.iop.org/0953-8984/13/i=9/a=312} {\bibfield  {journal}
  {\bibinfo  {journal} {J. Phys. Condens. Matter}\ }\textbf {\bibinfo {volume}
  {13}},\ \bibinfo {pages} {1891} (\bibinfo {year} {2001})}\BibitemShut
  {NoStop}%
\bibitem [{\citenamefont {Mirebeau}\ \emph {et~al.}(2005)\citenamefont
  {Mirebeau}, \citenamefont {Apetrei}, \citenamefont {Rodr\'{\i}guez-Carvajal},
  \citenamefont {Bonville}, \citenamefont {Forget}, \citenamefont {Colson},
  \citenamefont {Glazkov}, \citenamefont {Sanchez}, \citenamefont {Isnard},\
  and\ \citenamefont {Suard}}]{mirebeau2005ordered}%
  \BibitemOpen
  \bibfield  {author} {\bibinfo {author} {\bibfnamefont {I.}~\bibnamefont
  {Mirebeau}}, \bibinfo {author} {\bibfnamefont {A.}~\bibnamefont {Apetrei}},
  \bibinfo {author} {\bibfnamefont {J.}~\bibnamefont
  {Rodr\'{\i}guez-Carvajal}}, \bibinfo {author} {\bibfnamefont
  {P.}~\bibnamefont {Bonville}}, \bibinfo {author} {\bibfnamefont
  {A.}~\bibnamefont {Forget}}, \bibinfo {author} {\bibfnamefont
  {D.}~\bibnamefont {Colson}}, \bibinfo {author} {\bibfnamefont
  {V.}~\bibnamefont {Glazkov}}, \bibinfo {author} {\bibfnamefont {J.~P.}\
  \bibnamefont {Sanchez}}, \bibinfo {author} {\bibfnamefont {O.}~\bibnamefont
  {Isnard}}, \ and\ \bibinfo {author} {\bibfnamefont {E.}~\bibnamefont
  {Suard}},\ }\href {\doibase 10.1103/PhysRevLett.94.246402} {\bibfield
  {journal} {\bibinfo  {journal} {Phys. Rev. Lett.}\ }\textbf {\bibinfo
  {volume} {94}},\ \bibinfo {pages} {246402} (\bibinfo {year}
  {2005})}\BibitemShut {NoStop}%
\bibitem [{\citenamefont {Aleksandrov}\ \emph {et~al.}(1986)\citenamefont
  {Aleksandrov}, \citenamefont {Lidskii}, \citenamefont {Mamsurova},
  \citenamefont {Neigauz}, \citenamefont {Pigalskii}, \citenamefont {Pukhov},
  \citenamefont {Trusevich},\ and\ \citenamefont
  {Shcherbakova}}]{aleksandrov1986crystal}%
  \BibitemOpen
  \bibfield  {author} {\bibinfo {author} {\bibfnamefont {I.~V.}\ \bibnamefont
  {Aleksandrov}}, \bibinfo {author} {\bibfnamefont {B.~V.}\ \bibnamefont
  {Lidskii}}, \bibinfo {author} {\bibfnamefont {L.~G.}\ \bibnamefont
  {Mamsurova}}, \bibinfo {author} {\bibfnamefont {M.~G.}\ \bibnamefont
  {Neigauz}}, \bibinfo {author} {\bibfnamefont {K.~S.}\ \bibnamefont
  {Pigalskii}}, \bibinfo {author} {\bibfnamefont {K.~K.}\ \bibnamefont
  {Pukhov}}, \bibinfo {author} {\bibfnamefont {N.~G.}\ \bibnamefont
  {Trusevich}}, \ and\ \bibinfo {author} {\bibfnamefont {L.~G.}\ \bibnamefont
  {Shcherbakova}},\ }\href@noop {} {\bibfield  {journal} {\bibinfo  {journal}
  {JETP}\ }\textbf {\bibinfo {volume} {62}},\ \bibinfo {pages} {1287} (\bibinfo
  {year} {1986})}\BibitemShut {NoStop}%
\bibitem [{don()}]{donnerer2018dataset}%
  \BibitemOpen
  \href {http://dx.doi.org/10.14324/000.ds.10042850} {\bibinfo  {journal}
  {http://dx.doi.org/10.14324/000.ds.10042850}\ }\BibitemShut {NoStop}%
\end{thebibliography}
\end{document}